\documentclass[a4paper,UKenglish,cleveref, autoref, thm-restate]{lipics-v2021b}

\title{Probabilistic Model Checking for \\ Strategic Equilibria-based Decision Making: \\ Advances and Challenges}

\titlerunning{Probabilistic Model Checking for Strategic Equilibria-based Decision Making}

\author{Marta Kwiatkowska}{University of Oxford, Oxford, UK}{marta.kwiatkowska@cs.ox.ac.uk}{0000-0001-9022-7599}{}

\author{Gethin Norman}{University of Glasgow, Glasgow, UK \and University of Oxford, Oxford, UK}{gethin.norman@glasgow.ac.uk}{0000-0001-9326-4344}{}

\author{David Parker}{University of Birmingham, Birmingham, UK}{d.a.parker@cs.bham.ac.uk}{0000-0003-4137-8862}{}

\author{Gabriel Santos}{University of Oxford, Oxford, UK}{gabriel.santos@cs.ox.ac.uk}{0000-0002-6570-9737}{}

\author{Rui Yan}{University of Oxford, Oxford, UK}{rui.yan@cs.ox.ac.uk}{0000-0002-8685-5055}{}

\authorrunning{Marta Kwiatkowska et. al.} 

\Copyright{Marta Kwiatkowska and Gethin Norman and David Parker and Gabriel Santos and Rui Yan} 

\ccsdesc[10003752]{Theory of computation} 


\category{Invited Talk} 

\relatedversion{} 

\funding{This project was funded by the ERC under the European Union’s Horizon 2020 research and innovation programme (\href{http://www.fun2model.org}{FUN2MODEL}, grant agreement No.~834115).}

\usepackage{mathtools}
\usepackage{tikz}
\usetikzlibrary{arrows.meta}
\usetikzlibrary{arrows,automata}
\usetikzlibrary{shapes,decorations,patterns,positioning}
\usepackage{aircraftshapes}
\usepackage{pgfplots}
\pgfplotsset{compat=1.17} 

\usepackage{trimclip}
\usepackage{kbordermatrix}
\renewcommand{\kbldelim}{(}
\renewcommand{\kbrdelim}{)}

\definecolor{OliveGreen}{rgb}{0,0.6,0}
\definecolor{LimeGreen}{rgb}{0.2, 0.8, 0.2}

\DeclareMathOperator{\opt}{opt}
\renewcommand{\emptyset}{\varnothing}

\usepackage{scalerel}

\newcommand\scale[2]{\vstretch{#1}{\hstretch{#1}{#2}}}

\newcommand{\sectref}[1]{Section~\ref{#1}}

\newcommand{\figref}[1]{Figure~\ref{#1}}

\newcommand{\egref}[1]{Example~\ref{#1}}

\newcommand{\thmref}[1]{Theorem~\ref{#1}}

\newcommand{\defref}[1]{Definition~\ref{#1}}

\newcounter{exampcount}
\def\ra{{\rightarrow}}

\def\cC{{\mathcal{C}}}

\def\Nset{\mathbb{N}}
\def\Qset{\mathbb{Q}}
\def\Rset{\mathbb{R}}

\def\Eset{\mathbb{E}}

\def\ra{\rightarrow} 
\def\rmdef{\,\stackrel{\mbox{\rm {\tiny def}}}{=}}
\newcommand{\sem}[1]{ [ \! [ {#1} ] \! ]} 
\newcommand{\true}{\mathtt{true}} 

\renewcommand{\leq}{\leqslant}
\renewcommand{\geq}{\geqslant}

\def\squareforqed{\hbox{\rlap{$\sqcap$}$\sqcup$}}
\def\qed{\ifmmode\squareforqed\else{\unskip\nobreak\hfil
\penalty50\hskip1em\null\nobreak\hfil\squareforqed
\parfillskip=0pt\finalhyphendemerits=0\endgraf}\fi}

\newcommand\game{\textsf{G}}
\newcommand\nfgame{\textsf{N}}
\newcommand\mgame{\textsf{Z}}

\newcommand{\nscsg}{\mathsf{NSC}}
\newcommand{\agent}{\mathsf{Ag}}

\newcommand{\obs}{\mathit{obs}}

\newcommand{\Loc}{\mathit{Loc}}
\newcommand{\Per}{\mathit{Per}}

\newcommand{\loc}{\mathit{loc}}
\newcommand{\per}{\mathit{per}}

\newcommand\sinit{{\bar{s}}}

\newcommand\dist{{\mathit{Dist}}}

\newcommand\Prob{{\mathit{Prob}}}

\newcommand\val{{\mathit{val}}}

\newcommand{\last}{\mathit{last}}
\newcommand{\ipaths}{\mathit{IPaths}}
\newcommand{\fpaths}{\mathit{FPaths}}

\def\AP{{\mathit{AP}}}

\newcommand{\coalition}[1]{\langle \! \langle {#1} \rangle \! \rangle}

\def\next{{X\,}}
\def\until{{\ {\cal U}\ }}
\def\buntil{{\ {\cal U}^{\leq k}\ }}

\def\next{{\mathtt X\,}}
\def\until{{\ \mathtt{U}\ }}

\def\buntil{{\ \mathtt{U}^{\leq k}\ }}

\def\future{{\mathtt{F}\ }}
\def\futureop{{\mathtt{F}}}

\newcommand\bfuturep[1]{{\mathtt{F}^{\leq #1}\ }}

\newcommand{\scumul}[1]{\mathtt{C}^{#1}} 

\newcommand{\ap}{\mathsf{a}}
\newcommand{\sinstant}[1]{\mathtt{I}^{#1}} 

\newcommand\probopP{{\mathtt P}}
\newcommand\nashop[4]{\coalition{#1}({#2})_{#3}(#4)}

\newcommand\NE{\mbox{\rm \textsc{ne}}}
\newcommand\CE{\mbox{\rm \textsc{ce}}}
\newcommand\SW{\mbox{\rm \textsc{sw}}}
\newcommand\SF{\mbox{\rm \textsc{sf}}}

\newcommand\probop[2]{\probopP_{#1}[\,{#2}\,]}

\newcommand\rewopR{{\mathtt R}}
\newcommand\rewop[3]{\rewopR^{#1}_{#2}[\,{#3}\,]}

\newcommand{\lab}{{\mathit{L}}}

\newcommand{\marta}[1]{\marginpar{\footnotesize \color{purple} {\bf M:} \textsf{#1}}}

\newcommand{\gabriel}[1]{\marginpar{\footnotesize \color{orange} {\bf GS:} \textsf{#1}}}

\EventEditors{Stefan Szeider, Robert Ganian, and Alexandra Silva}
\EventNoEds{3}
\EventLongTitle{47th International Symposium on Mathematical Foundations of Computer Science (MFCS 2022)}
\EventShortTitle{MFCS 2022}
\EventAcronym{MFCS}
\EventYear{2022}
\EventDate{August 22--26, 2022}
\EventLocation{Vienna, Austria}
\EventLogo{}
\SeriesVolume{241}
\ArticleNo{4}
\begin{document}

\maketitle

\begin{abstract}
Game-theoretic concepts have been extensively studied in economics to provide insight into competitive behaviour and strategic decision making. As computing systems increasingly involve concurrently acting autonomous agents, game-theoretic approaches are becoming widespread in computer science as a faithful modelling abstraction. These techniques can be used to reason about the competitive or collaborative behaviour of multiple rational agents with distinct goals or objectives. This paper provides an overview of recent advances in developing a modelling, verification and strategy synthesis framework for concurrent stochastic games implemented in the probabilistic model checker PRISM-games. This is based on a temporal logic that supports finite- and infinite-horizon temporal properties in both a zero-sum and nonzero-sum setting, the latter using Nash and correlated equilibria with respect to two optimality criteria, social welfare and social fairness. We summarise the key concepts, logics and algorithms and the currently available tool support. Future challenges and recent progress in adapting the framework and algorithmic solutions to continuous environments and neural networks are also outlined.       
\end{abstract}

\section{Introduction}\label{intro-sect}

Game-theoretic techniques have long been a source of fundamental insights into strategic decision making for multi-agent systems. They have been widely studied in areas such as economics \cite{HR06}, control \cite{MS18} and robotics \cite{LaV00}.  Concurrent stochastic multi-player games (CSGs), in particular, provide a natural framework for modelling a set of interactive, rational agents operating concurrently
within an uncertain or stochastic environment. 
They can be viewed as a collection of players (agents) with
strategies for determining their actions based on the execution so far,
and where the resulting evolution of the system is probabilistic.

Game-theoretic analysis is versatile, in that it can support both zero-sum and nonzero-sum (equilibria) analysis. Zero-sum properties focus on scenarios in
which one player (or a coalition of players) aims to optimise some objective, while
the remaining players form a coalition with the directly opposing goal. On the other hand, nonzero-sum (equilibria) properties correspond to situations
where two or more players (or coalitions of players) in a CSG have distinct objectives
to be maximised or minimised. In nonzero-sum properties the goals of the players (or coalitions) are not
necessarily directly opposing, and therefore it may be beneficial for players to collaborate.
Competitive scenarios occur in many applications, e.g., attackers and defenders in the context of computer security.
Similarly, collaborative behaviour can be essential, e.g., to effectively control a multi-robot system,
or for users to send data efficiently through a shared medium in a communication protocol.

Probabilistic model checking is a powerful approach to the formal analysis
of systems with stochastic behaviour. It relies on the construction and analysis of a probabilistic model,
guided by a formal specification of its desired behaviour in temporal logic.
It is of particular benefit in the context of models, such as stochastic games,
which combine nondeterministic and probabilistic behaviour.
This is because the interplay between these aspects of the model can be subtle
and lead to unexpected results if not carefully modelled and analysed.
This is exacerbated when the system comprises multiple agents with differing objectives.

Until recently, practical applications of probabilistic model checking based on stochastic games had focused primarily on \emph{turn-based} models~\cite{CFK+13b}, in which simultaneous decision making by agents is forbidden.
Alternatively, model checking of \emph{non-stochastic} games has been
extensively studied, and tool support developed~\cite{AHM+98,LQR09}.
CSGs provide a more powerful and realistic modelling formalism,
but also bring considerable challenges, in terms of the higher 
computational complexity or undecidability for some key problems.

There has nonetheless been significant amounts of work on tackling verification problems for CSGs.
A number of algorithms have been proposed for solving CSGs against formally specified zero-sum properties,
e.g.~\cite{AHK07,AM04,CAH13}.
In the case of nonzero-sum properties, \cite{CMJ04,GLH+21} 
study the existence of and the complexity of finding equilibria for stochastic games. Complexity results for finding equilibria are also considered in \cite{BBG+19} and \cite{GNP+19} for quantitative reachability properties and temporal logic properties, respectively.
Other work concerns finding equilibria for discounted properties; we mention \cite{PPB15}, which formulates a learning-based algorithm, and \cite{LP17}, which presents iterative algorithms. 
However these advances are mostly lacking in implementations, tool support or case studies. 
Tools exist for
solving turn-based stochastic games~\cite{CHJ+10,CKL+11}
and non-stochastic concurrent games~\cite{CKL+11,BRE13,CLM+14,TGW15,GNP+18,GAMB},
with the latter class including support for computing equilibria.



At the same time, there is an increasing trend to incorporate data-driven decision making, which necessitates the incorporation on of machine learning components within autonomous systems, which are built largely using conventional, symbolic methods. Examples of such \emph{neuro-symbolic} systems are self-driving cars whose vision function is provided via a neural network image classifier, or an aircraft controller whose collision avoidance system uses a neural network for decision support. 
Design automation support for such systems is lacking, yet automatic computation of equilibria aids in ensuring stable solutions. 

This paper provides an overview of recent advances in developing a modelling, verification and strategy synthesis framework for concurrent stochastic games, as implemented in the PRISM-games probabilistic model checker~\cite{KNPS20}. The framework uses a temporal logic that supports a wide range of finite- and infinite-horizon properties,
relating to the probability of an event's occurrence or the expected amount of reward or cost accumulated.
The logic allows specification of both \emph{zero-sum} and \emph{nonzero-sum} properties,
with the latter expressed using either \emph{Nash equilibria} or \emph{correlated equilibria}.
For both types of equilibria, strategies are synthesised in which it is not beneficial for any player to unilaterally alter their chosen actions, but correlated equilibria also allow players to coordinate through \emph{public signals}.
Since several, varied such equilibria may exist, we also support distinct optimality criteria to select between them; we consider \emph{social welfare}, which maximises the sum of the players utilities, and \emph{social fairness}, which minimises the difference between the utilities.

We summarise the key concepts, logics and algorithms that underlie this framework
and discuss the tool support provided by PRISM-games, including an illustrative case study
of formally modelling and analysing a multi-agent communication protocols using CSGs.
Future challenges and recent progress in extending the framework and algorithmic solutions to modelling of neuro-symbolic CSGs are also outlined. In contrast to the majority of prior research, the focus of this strand of work is on software tool development, applications and case studies. 

\section{Normal form games}\label{nfgs-sect}

We introduce the main concepts used in this paper by means of simple one-shot games known as \emph{normal form games} (NFGs), where players make their choices at the same time.
We consider both zero-sum NFGs and nonzero-sum NFGs, then define equilibria concepts for these games and summarise existing algorithms for equilibria computation. 

We first require the following notation. Let $\dist(X)$ denote the set of probability distributions over set $X$. For any vector $v \in \Rset^n$, we use $v(i)$ to refer to the $i$th entry of the vector. For any tuple $x=(x_1,\dots,x_n) \in X^n$, element $x' \in X$ and $i \leq n$, we define the tuples $x_{-i} \rmdef (x_1,\dots,x_{i-1},x_{i+1},\dots,x_n)$ and $x_{-i}[x'] \rmdef (x_1,\dots,x_{i-1},x',x_{i+1},\dots,x_n)$.

\begin{definition}[Normal form game] 
A (finite, $n$-player) \emph{normal form game} (NFG) is a tuple $\nfgame = (N,A,u)$ where: 
\begin{itemize}
\item
$N=\{1,\dots,n\}$ is a finite set of \emph{players};
\item
$A = A_1 {\times} \cdots {\times} A_n$ and $A_i$ is a finite set of \emph{actions} available to player $i \in N$;
\item
$u = (u_1,\dots,u_n)$ and $u_i \colon A \rightarrow \Rset$ is a \emph{utility function} for player $i \in N$.
\end{itemize}
\end{definition}

\noindent
In a normal form game $\nfgame$, the players choose actions simultaneously, with player $i \in N$ choosing an action from the set $A_i$ and, assuming that each player $i \in N$ selects action $a_i$, player $j$ receives the utility $u_j(a_1,\dots,a_n)$. The \emph{objective} of each player is to maximise their utility and their choices are governed by \emph{strategies}, which we now define. We will also distinguish \emph{strategy profiles}, which comprise a strategy for each player, and \emph{correlated profiles}, which correspond to choices of the players when they are allowed to coordinate through a (probabilistic) \emph{public signal}.
\begin{definition}[Strategies, profiles and correlated profiles]\label{strats-nfgs}
For an NFG $\nfgame$:
\begin{itemize}
\item
a \emph{strategy} $\sigma_i$ for player $i$ in an NFG $\nfgame$ is a probability distribution over the set of actions $A_i$ and we let $\Sigma^i_\nfgame$ denote the set of all strategies for player $i$;
\item
a \emph{strategy profile} (or \emph{profile})
$\sigma = (\sigma_1,\dots,\sigma_n)$ is a tuple of strategies for each player;
\item 
a \emph{correlated profile} is a tuple $(\tau,\varsigma)$ comprising $\tau \in \dist(D_1{\times}\cdots{\times}D_n)$, where $D_i$ is a finite set of \emph{signals} for player $i$,
and $\varsigma=(\varsigma_1,\dots,\varsigma_n)$, where $\varsigma_i \colon D_i \ra A_i$ is a function from the signals of player $i$ to the actions of player $i$.
\end{itemize}
\end{definition}

\noindent
For a correlated profile $(\tau,\varsigma)$ of $\nfgame$, the public signal $\tau$ is a joint distribution over signals $D_i$ for each player $i$ such that, if player $i$ receives the signal $d_i \in D_i$, then it chooses action $\varsigma_i(d_i)$. 
We can consider any correlated profile $(\tau,\varsigma)$ as a \emph{joint strategy}, i.e., a distribution over $A_1 {\times}\cdots{\times}A_n$ where:
\[ \begin{array}{c}
(\tau,\varsigma)(a_1,\dots,a_n) = \sum \{ \tau(d_1,\dots,d_n) \mid d_i \in D_i \wedge \varsigma(d_i)=a_i \; \mbox{for all $i \in N$} \} \, .
\end{array}
\]
Conversely, any joint strategy $\tau \in \dist(A_1 {\times}\cdots{\times}A_n)$ of $\nfgame$ can be considered as a correlated profile $(\tau,\varsigma)$, where $D_i=A_i$ and $\varsigma_i$ is the identity function for $i \in N$.
Any profile $\sigma$ of an NFG $\nfgame$ can be mapped to an equivalent correlated profile (in which $\tau$ is the joint distribution $\sigma_1 {\times}\cdots {\times} \sigma_n$ and $\varsigma_i$ is the identity function). On the other hand, there are correlated profiles with no equivalent strategy profile.

Under profile $\sigma$ or correlated profile $(\tau,\varsigma)$ the expected utilities of player $i$ are:
\[ 
\begin{array}{rcl}
u_i(\sigma) & \rmdef & \sum_{(a_1,\dots,a_n) \in A} u_i(a_1,\dots,a_n) \cdot \big( \prod_{j=1}^n \sigma_j(a_j) \big) \\
u_i(\tau,\varsigma) & \rmdef & \sum_{(d_1,\dots,d_n) \in D} \tau(d_1,\dots,d_n) \cdot u_i(\varsigma_1(d_1),\dots,\varsigma_n(d_n)) \, .
\end{array} 
\]
\begin{example}
Consider the two-player NFG with available action sets $A_i = \{ \mathit{heads}_i , \mathit{tails}_i \}$ for $1 \leq i \leq 2$ and a correlated profile corresponding to the joint distribution $\tau \in \dist(A_1{\times}A_2)$, where $\tau(\mathit{heads}_1,\mathit{heads}_2) = \tau(\mathit{tails}_1,\mathit{tails}_2) = 0.5$.
Under this correlated profile, the players share a fair coin and choose their action based on the outcome of the coin toss. There is no equivalent strategy profile.
\end{example}

\subsection{Zero-sum NFGs}\label{zerocsgs-sect}

A \emph{zero-sum NFG} is a two-player NFG $\nfgame$ such that $u_1(\alpha) {+} u_2(\alpha) = 0$ for all $\alpha \in A$, meaning that 
the objectives of the players are directly opposing.  Such an NFG is often called a \emph{matrix game}, as it can be represented by a single matrix $\mgame \in \Qset^{l \times m}$, where $A_1 = \{a_1,\dots,a_l\}$, $A_2 = \{b_1,\dots,b_m\}$ and $z_{ij} = u_1(a_i,b_j) = - u_2(a_i,b_j)$. 

We next introduce the notion of the \emph{value} of a zero-sum NFG and recall classical results about the existence of optimal strategies. 
%
\begin{theorem}[Minimax theorem~\cite{NEU28,NMK+44}]\label{minimax-thm}
For any zero-sum NFG $\nfgame = (N,A,u)$ and corresponding matrix game $\mgame$, there exists $v^\star \in \Qset$, called the \emph{value} of the game and denoted $\val(\mgame)$, such that:
\begin{itemize}
\item there is a strategy $\sigma_1^\star$ for player 1, called an optimal strategy of player 1, such that under this strategy the player's expected utility is at least $v^\star$ regardless of the strategy of player 2, i.e., $\inf_{\sigma_2 \in \Sigma^2_\nfgame} u_1(\sigma_1^\star,\sigma_2) \geq v^\star$;
\item there is a strategy $\sigma_2^\star$ for player 2, called an optimal strategy of player 2, such that under this strategy the player's expected utility is at least $-v^\star$ regardless of the strategy of player 1, i.e., $\inf_{\sigma_1 \in \Sigma^1_\nfgame} u_2(\sigma_1,\sigma_2^\star) \geq -v^\star$.
\end{itemize}
\end{theorem}

\noindent The value of a matrix game $\mgame \in \Qset^{l \times m}$ can be found by solving a linear programming (LP) problem~\cite{NEU28,NMK+44}.

\begin{example} Table \ref{tab:matching_pennies_nfg} shows a classic example of a two-player zero-sum game known as \emph{matching pennies}. Columns $\alpha$ and $u_i$ represent the collective choice (profile) and player $i$'s utility, respectively. In this example, each player has a coin for which they may choose the value to be heads or tails, i.e., $A_i=\{\mathit{heads}_i, \mathit{tails}_i\}$. If the coins match, player 1 wins the round, which is indicated by being awarded a utility of $1$, while player 2 receives utility $-1$. If the coins do not match, then the players' utilities are negated.

\begin{table}[h]
\centering
\begin{tabular}{c}
\begin{tabular}{ccc}
$\alpha$                                 & $u_1(\alpha)$                  & $u_2(\alpha)$                     \\ \hline
\multicolumn{1}{|c|}{$(\mathit{heads}_1, \mathit{heads}_2)$} & \multicolumn{1}{r|}{$1$} & \multicolumn{1}{r|}{$-1$} \\ \hline
\multicolumn{1}{|c|}{$(\mathit{heads}_1, \mathit{tails}_2)$} & \multicolumn{1}{r|}{$-1$} & \multicolumn{1}{r|}{$1$} \\ \hline
\end{tabular}
\begin{tabular}{ccc}
$\alpha$                                 & $u_1(\alpha)$                  & $u_2(\alpha)$                     \\ \hline
\multicolumn{1}{|c|}{$(\mathit{tails}_1, \mathit{heads}_2)$} & \multicolumn{1}{r|}{$-1$} & \multicolumn{1}{r|}{$1$} \\ \hline
\multicolumn{1}{|c|}{$(\mathit{tails}_1, \mathit{tails}_2)$} & \multicolumn{1}{r|}{$1$} & \multicolumn{1}{r|}{$-1$} \\ \hline
\end{tabular}
\end{tabular}
\vspace{0.3cm}
\caption{Matching pennies game in normal form.}
\label{tab:matching_pennies_nfg}
\vspace{-0.3cm}
\end{table}
The value for the corresponding matrix game is the solution to the following LP problem: Maximise $v$ subject to:
\begin{eqnarray*}
x_1 - x_2 \geq v,\ 
x_2 - x_1 \geq v,\ 
x_1 + x_2 = 1
\end{eqnarray*}
which yields the value $v^\star=0$ with optimal strategy $\sigma_1^\star = (\frac{1}{2},\frac{1}{2})$ for player 1 (the optimal strategy for player 2 is the same).
\end{example}

\subsection{Nonzero-sum NFGs} 
The requirement for players to have directly opposing objectives is often too limiting, and it is necessary to allow distinct objectives, which cannot be modelled in a zero-sum fashion. These scenarios can be captured using the notion of \emph{equilibria}, defined by a separate, independent objective for each agent. We now define the concepts of \emph{Nash equilibrium}~\cite{NMK+44} and \emph{correlated equilibrium}~\cite{Aum74} for NFGs, which ensure stability against deviations by individual agents, improving the overall game outcomes. Since many equilibria may exist, we also introduce optimality criteria for these equilibria: \emph{social welfare}, which is standard \cite{NRTV07}, and \emph{social fairness}, which was first defined in \cite{KNPS22}. 

Before giving the formal definitions, we first extend our notation as follows: for any profile $\sigma$ and strategy $\sigma^\star_i$, the strategy tuple $\sigma_{-i}$ corresponds to $\sigma$ with the strategy of player $i$ removed and $\sigma_{-i}[\sigma^\star_i]$ to the profile $\sigma$ after replacing player $i$'s strategy with $\sigma^\star_i$.
\begin{definition}[Best response]\label{def:bestresponse} 
For any nonzero-sum NFG $\nfgame$ and profile $\sigma$ or correlated profile $(\tau,\varsigma)$ of $\nfgame$, 
the \emph{best response} moves for player $i$ to $\sigma_{-i}$ and $(\tau,\varsigma_{-i})$ are, respectively:
\begin{itemize}
\item
a strategy $\sigma^\star_i$ for player $i$ such that $u_i(\sigma_{-i}[\sigma^\star_i]) \geq u_i(\sigma_{-i}[\sigma_i])$ for all $\sigma_i \in \Sigma^i_\nfgame$;
\item
a function $\varsigma^\star_i \colon D_i \ra A_i$ for player $i$ such that $u_i(\tau,\varsigma_{-i}[\varsigma^\star_i]) \geq u_i(\tau,\varsigma_{-i}[\varsigma_i])$ for all functions $\varsigma_i \colon D_i \ra A_i$.
\end{itemize}
\end{definition}

\begin{definition}[NE and CE]\label{def:eq}
For any nonzero-sum NFG $\nfgame$, a strategy profile $\sigma^\star$ is a \emph{Nash equilibrium} (NE) and a correlated profile $(\tau,\varsigma^\star)$ of $\nfgame$ is a \emph{correlated equilibrium} (CE) if:
\begin{itemize}
\item
$\sigma_i^\star$ is a best response to $\sigma_{-i}^\star$ for all $i \in N$;
\item
$\varsigma_i^\star$ is a best response to $(\tau,\varsigma_{-i}^\star)$ for all $i \in N$;
\end{itemize}
respectively.
\end{definition}

\noindent
Any NE of $\nfgame$ is also a CE, while there exist CEs that cannot be represented by a strategy profile, and therefore are not NEs. 
For each class of equilibria, NE and CE, we introduce two optimality criteria, the first maximising \emph{social welfare} (SW), defined as the \emph{sum} of the utilities, and the second maximising \emph{social fairness} (SF), which minimises the \emph{difference} between the players' utilities. Other variants of fairness have been considered for NEs, such as in \cite{LRTZ06}, where the authors seek to maximise the lowest utility among the players.
\begin{definition}[SW and SF]\label{def:swne}
An equilibrium $\sigma^\star$ is a \emph{social welfare} (SW) equilibrium if the sum of the utilities of the players under $\sigma^\star$ is maximal over all equilibria, while $\sigma^\star$ is a \emph{social fair} (SF) equilibrium if the difference between the player's utilities under $\sigma^\star$ is minimised over all equilibria.
\end{definition}

\noindent
We can also define the dual concept of \emph{social cost} (SC) equilibria~\cite{KNPS21}, where players try to minimise, rather than maximise, their expected utilities by considering equilibria of the game $\nfgame^{-}= (N,A,{-}u)$ in which the utilities of $\nfgame$ are negated. We remark that SC equilibria strategies are not a subset of classically defined NE or CE strategies of $\nfgame$.
\def\xmin{0}%
\def\xmax{6}%
\def\ymin{0}%
\def\ymax{6}%

\def\xra{2}
\def\yra{5}
\def\xrb{0}
\def\yrb{2}
\def\xrc{3}
\def\yrc{0}

\begin{figure}[t]
\begin{subfigure}{0.41\textwidth}
\centering
\begin{tikzpicture}[scale=0.6]
    



\draw[OliveGreen, very thick, fill=LimeGreen, opacity=0.8] (0,0) rectangle (1.5,1.5);
\draw[OliveGreen, very thick, fill=LimeGreen, opacity=0.8] (4.5,4.5) rectangle (6,6);
\draw[OliveGreen, very thick, fill=LimeGreen, opacity=0.8] (0,4.5) rectangle (1.5,6);
\draw[OliveGreen, very thick, fill=LimeGreen, opacity=0.8] (4.5,0) rectangle (6,1.5);






\draw[fill=gray, opacity=0.2] (1.5,0) rectangle (4.5,1.5);
\draw[fill=gray, opacity=0.2] (0,1.5) rectangle (1.5,4.5);
\draw[fill=gray, opacity=0.2] (1.5,1.5) rectangle (4.5,4.5);
\draw[fill=gray, opacity=0.2] (1.5,4.5) rectangle (4.5,6);
\draw[fill=gray, opacity=0.2] (4.5,1.5) rectangle (6,4.5);

\draw[-,white,thick,dashed] (0.0,3.0) -- (6.0,3.0);
\draw[-,white,thick,dashed] (3.0,0.0) -- (3.0,6.0);

\draw[->,blue,thick,dashed] (1.35,2.25) -- (6.0,2.25);

\draw[->,orange,thick,dashed] (2.25,4.7) -- (2.25,0.0);
\draw[->,red,thick,dashed] (3.75,1.3) -- (3.75,6.0);



\node[fill, color=orange, opacity=0.7, rectangle, draw, radius=0.2, inner sep=6pt, label={}, scale=0.6] at (\xra+0.25,\yra+0.225) (ra) {\Large \color{white} \textbf{$\mathbf{c_1}$}};
\node[fill, color=blue, opacity=0.7, rectangle, draw, radius=0.2, inner sep=6pt, label={}, scale=0.6] at (\xrb+0.7,\yrb+0.25) (rb) {\Large \color{white} \textbf{$\mathbf{c_2}$}};
\node[fill, color=red, opacity=0.7, rectangle, draw, radius=0.2, inner sep=6pt, label={}, scale=0.6] at (\xrc+0.75,\yrc+0.75) (rc) {\Large \color{white} $\mathbf{c_3}$};



\end{tikzpicture}
\vspace*{-0.4cm}
\end{subfigure}
\centering
\begin{subfigure}{0.58\textwidth}
\begin{tabular}{cccc}
$\alpha$                                 & $u_1(\alpha)$                  & $u_2(\alpha)$                  & $u_3(\alpha)$                  \\ \hline
\multicolumn{1}{|c|}{$({\textit{pro}}_1, {\textit{pro}}_2, {\textit{pro}}_3)$} & \multicolumn{1}{c|}{$-1000$} & \multicolumn{1}{c|}{$-1000$} & \multicolumn{1}{c|}{$-100$} \\ \hline
\multicolumn{1}{|c|}{$({\textit{pro}}_1, {\textit{pro}}_2, {\textit{yld}}_3)$} & \multicolumn{1}{c|}{$-1000$} & \multicolumn{1}{c|}{$-100$} & \multicolumn{1}{c|}{$-5$} \\ \hline
\multicolumn{1}{|c|}{$({\textit{pro}}_1, {\textit{yld}}_2, {\textit{pro}}_3)$} & \multicolumn{1}{c|}{$5$} & \multicolumn{1}{c|}{$-5$} & \multicolumn{1}{c|}{5} \\ \hline
\multicolumn{1}{|c|}{$({\textit{pro}}_1, {\textit{yld}}_2, {\textit{yld}}_3)$} & \multicolumn{1}{c|}{$5$} & \multicolumn{1}{c|}{$-5$} & \multicolumn{1}{c|}{$-5$} \\ \hline
\multicolumn{1}{|c|}{$({\textit{yld}}_1, {\textit{pro}}_2, {\textit{pro}}_3)$} & \multicolumn{1}{c|}{$-5$} & \multicolumn{1}{c|}{$-1000$} & \multicolumn{1}{c|}{$-100$} \\ \hline
\multicolumn{1}{|c|}{$({\textit{yld}}_1, {\textit{pro}}_2, {\textit{yld}}_3)$} & \multicolumn{1}{c|}{$-5$} & \multicolumn{1}{c|}{$5$} & \multicolumn{1}{c|}{$-5$} \\ \hline
\multicolumn{1}{|c|}{$({\textit{yld}}_1, {\textit{yld}}_2, {\textit{pro}}_3)$} & \multicolumn{1}{c|}{$-5$} & \multicolumn{1}{c|}{$-5$} & \multicolumn{1}{c|}{$5$} \\ \hline
\multicolumn{1}{|c|}{$({\textit{yld}}_1, {\textit{yld}}_2, {\textit{yld}}_3)$} & \multicolumn{1}{c|}{$-10$} & \multicolumn{1}{c|}{$-10$} & \multicolumn{1}{c|}{$-10$} \\ \hline
\end{tabular}
\end{subfigure}
\vspace*{-0.2cm}
\caption{Example from \cite{KNPS22}: Cars at an intersection and the corresponding NFG.}\label{intersection:fig}
\vspace*{-0.2cm}
\end{figure}
\begin{example} Consider the scenario from \cite{KNPS22}, based on an example from~\cite{EP14}, where three cars meet at an intersection and want to proceed as indicated by the arrows in \figref{intersection:fig}. Each car can either \emph{proceed} or \emph{yield}. If two cars with intersecting paths proceed, then there is an accident. If an accident occurs, the car having the right of way, i.e., the other car is to its left,
has a utility of $-100$ and the car that should yield has a utility of $-1000$. If a car proceeds without causing an accident, then its utility is $5$ and the cars that  yield have a utility of $-5$. If all cars yield, then, since this delays all cars, all have utility $-10$. The 3-player NFG is given in~\figref{intersection:fig}.
The different optimal equilibria of the NFG are:
\begin{itemize}
\item
the SWNE and SWCE are the same: for $c_2$ to yield and $c_1$ and $c_3$ to proceed, with the expected utilities of the players $(5,-5,5)$;
\item
the SFNE is for $c_1$ to yield with probability $1$, $c_2$ to yield with probability $0.863636$ and $c_3$ to yield with probability $0.985148$, with the expected utilities of the players $(-9.254050, -9.925742, -9.318182)$; 
\item
the SFCE gives a joint distribution where the probability of $c_2$ yielding and of $c_1$ and $c_3$ yielding are both $0.5$ with the expected utilities of the players $(0,0,0)$. 
\end{itemize}
Modifying $u_2$ such that $u_2(\textit{pro}_1,\textit{pro}_2,\textit{pro}_3)=-4.5$ to, e.g., represent a reckless driver, the SWNE becomes for $c_1$ and $c_3$ to yield and $c_2$ to proceed with the expected utilities of the players $(-5,5,-5)$, while
the SWCE is still for $c_2$ to yield and $c_1$ and $c_3$ to proceed. 
The SFNE and SFCE also do not change.
\end{example}

\subparagraph*{Algorithms for computing equilibria in NFGs.} 

Finding 
NEs in two-player NFGs is in the class of \emph{linear complementarity} problems (LCPs). Established algorithms include the Lemke-Howson algorithm \cite{LH64}, which is based on the method of labelled polytopes \cite{NRTV07}, support enumeration \cite{PNS04} and regret minimisation \cite{SGC05}. In \cite{KNPS21} a method for NE computation is developed, which  reduces the problem to SMT via labelled polytopes~\cite{NRTV07} by considering the regions of the strategy profile space. This method iteratively reduces the search space of profiles as positive probability assignments are found and added as constraints on the profiles. This approach can also be used for finding both an SWNE and SFNE by computing all NEs and then selecting an optimal one.

In the case of NFGs with more than two players, the computation of NEs is more complex since,
for a given support (i.e., a sub-region of the strategy profile space which fixes the set of actions chosen with nonzero probability by each player), finding NEs cannot be reduced to an LP problem. A method for such NFGs is presented in~\cite{KNPS20b}, based on support enumeration~\cite{PNS04}, which exhaustively examines all supports one at a time, checking whether that sub-region contains NEs. For each support, finding an SWNE can be reduced to a \emph{nonlinear programming problem}~\cite{KNPS20b}. This nonlinear programming problem can be modified to find an SFNE in each support~\cite{KNPS22}.

In the case of CEs, the approach introduced in \cite{KNPS22} is to first find a joint strategy for the players, i.e., a distribution over the action tuples, which can then be mapped to a correlated profile. For SWCEs, \cite{KNPS22} reduces the computation to solving a LP problem which has $|A|$ variables, one for each action tuple, and $\sum_{i \in N}(|A_i|^2-|A_i|) + |A| + 1$ constraints. For SFCEs, on the other hand, the method of \cite{KNPS22} involves 
solving an optimisation problem with an additional has $|N|+2$ variables and $3\cdot|N|$ constraints compared to the LP problem for finding SWCEs.

\section{Concurrent Stochastic Games}\label{csgs-sec}

This section introduces \emph{concurrent stochastic games} (CSGs)~\cite{Sha53}, in which players repeatedly make simultaneous choices over actions and the action choices cause a probabilistic update of the game state. CSGs thus provide a natural framework for modelling a set of interactive, rational agents
operating concurrently within an uncertain or probabilistic environment.
Compared to normal form games, they are classified as \emph{multi-stage},
which is more convenient for specifying repeated or sequential interactions among agents.
The introduction of stochasticity facilitates modelling of a wide range of important phenomena,
for example uncertain behaviour due to noisy sensors or unreliable hardware in a multi-robot system,
or the use of randomisation for coordination in a distributed security or networking protocol.


\begin{definition}[Concurrent stochastic game]\label{csg-def} A \emph{concurrent stochastic multi-player game} (CSG) is a tuple
$\game = (N, S, \sinit, A, \Delta, \delta)$ where:
\begin{itemize}
\item $N=\{1,\dots,n\}$ is a finite set of players;
\item $S$ is a finite set of states and $\sinit \in S$ is an initial state;
\item $A = (A_1\cup\{\bot\}) {\times} \cdots {\times} (A_n\cup\{\bot\})$ where $A_i$ is a finite set of actions available to player $i \in N$ and $\bot$ is an idle action disjoint from the set $\cup_{i=1}^n A_i$;
\item $\Delta \colon S \rightarrow 2^{\cup_{i=1}^n A_i}$ is an action assignment function;
\item $\delta \colon S {\times} A \rightarrow \dist(S)$ is a probabilistic transition function.
\end{itemize}
\end{definition}

\noindent
Given a CSG $\game$, the set of actions available to player $i \in N$ in state $s \in S$ is given by $A_i(s) \rmdef \Delta(s) \cap A_i$. The CSG $\game$ starts in the initial state $\sinit$ and, if $\game$ is in the game state $s$, then  each player $i \in N$ selects an action from its available actions in state $s$ if this set is non-empty, and from $\{ \bot \}$ otherwise. Next, supposing each player $i \in N$ chooses action $a_i$, the game state is updated according to the distribution $\delta(s,(a_1,\dots,a_n))$.
We allow sets of players $C \subseteq N$ to form \emph{coalitions}, and will consider the induced CSG, called the \emph{coalition game}, with coalitions as players.

A \emph{path} $\pi$ of a CSG $\game$ is a sequence $\pi = s_0 \xrightarrow{\alpha_0}s_1 \xrightarrow{\alpha_1} \cdots$, 
where $s_i \in S$, $\alpha_i\in A$ 
and
$\delta(s_i,\alpha_i)(s_{i+1})>0$ for all $i \geq 0$. We denote by $\fpaths_{\game,s}$ and $\ipaths_{\game,s}$ the sets of finite and infinite paths starting in state $s$ of $\game$, respectively, and drop the subscript $s$ when considering all finite and infinite paths of $\game$. As for NFGs, we can define \emph{strategies} of $\game$ that resolve the choices of the players. Here, a strategy for player $i$ is a function $\sigma_i \colon \fpaths_{\game} \ra \dist(A_i \cup \{ \bot \})$ 
mapping finite paths to distributions over available actions,
such that, if $\sigma_i(\pi)(a_i){>}0$, then $a_i \in A_i(\last(\pi))$ where $\last(\pi)$ is the final state of $\pi$. Furthermore, we can define strategy profiles, correlated profiles and joint strategies analogously to \sectref{nfgs-sect}. 

A \emph{labelled} CSG is a tuple $(\game, \AP, \lab)$, where $\game$ is a CSG, as in \defref{csg-def}, $\AP$ is a set of atomic propositions and $\lab \colon S \rightarrow 2^{\AP}$ is a labelling function, specifying which atomic propositions are true in each state. We also associate CSGs with \emph{reward structures}, which annotate states and transitions with real values.
More precisely a reward structure is a pair $r{=}(r_A,r_S)$
consisting of an action reward function $r_A \colon S {\times} A \ra \Rset$ and state reward function $r_S \colon S \ra \Rset$.
We use atomic propositions and rewards as the building blocks to specify players' utilities in a CSG,
which will be described in \sectref{logic-sect}.

Formally, the utility function or \emph{objective} of player $i$ in a CSG is given by a random variable $X_i \colon \ipaths_{\game} \rightarrow \Rset$ over infinite paths. For a profile $\sigma$ and state $s$, using standard techniques~\cite{KSK76}, we can construct a probability measure $\Prob^{\sigma}_{\game,s}$ over the paths that start in state $s$ corresponding to $\sigma$, denoted $\ipaths^\sigma_{\game,s}$, and define the expected value $\Eset^{\sigma}_{\game,s}(X_i)$ of player $i$'s utility from $s$ under $\sigma$. Similarly, we can also define such a probability measure and expected value given a correlated profile or joint strategy of $\game$.

\subsection{Zero-sum CSGs}

Similarly to NFGs (see \sectref{zerocsgs-sect}),  \emph{zero-sum} CSGs are two-player games that have a single utility function $X$ for player 1, with the utility function of player 2 given by $-X$, and both players aiming to maximise the expected value of their utility. Equivalently, we can suppose that player 1 tries to maximise the expected value of $X$, while player 2 tries to minimise it. As for NFGs (see \thmref{minimax-thm}), a CSG has a \emph{value} with respect to $X$ if it is determined, i.e., if the maximum value that player 1 can ensure equals the minimum value that player 2 can ensure when starting from any state of the CSG. Since the CSGs we discuss in this paper are finite-state and finitely-branching, it follows that they are determined for all of the objectives that we consider~\cite{Mar98}.

Given a multi-player CSG and objective $X$,
we can divide the players into two coalitions, $C\subseteq N$ and $N\backslash C$,
and then construct a \emph{two-player} zero-sum coalition game, in which each coalition acts as a single player, with one coalition trying to maximise the value of $X$ and the other trying to minimise that value.

\subsection{Nonzero-sum CSGs} 

We define \emph{nonzero-sum} CSGs similarly to NFGs: we assume that there is a distinct and independent objective $X_i$ for each player $i$ (or coalitions of players). We can then define NE and CE for CSGs (see \defref{def:eq}), as well as the restricted classes of SW and SF equilibria, similarly to those for NFGs (see \defref{def:swne}).
Following \cite{KNPS21,KNPS20b}, we focus on \emph{subgame-perfect} equilibria~\cite{OR04}, which are equilibria in \emph{every state} of $\game$. Furthermore, because we include infinite-horizon objectives,
where the existence of NE is an open problem~\cite{BMS14},
we will in some cases use $\varepsilon$-NE, which do exist for any $\varepsilon>0$ for all the infinite-horizon objectives we consider.
%
\begin{definition}[Subgame-perfect $\varepsilon$-NE] For CSG $\game$ and $\varepsilon>0$, a strategy profile $\sigma^\star$ is a \emph{subgame-perfect $\varepsilon$-Nash equilibrium}
for objectives $\langle X_i \rangle_{i \in N}$ if and only if\/ $\Eset^{\sigma^\star}_{\game,s}(X_i) \geq \sup_{\sigma_i \in \Sigma_i} \Eset^{\sigma^\star_{-i}[\sigma_i]}_{\game,s}(X_i) - \varepsilon$ for all $i \in N$ and $s \in S$.
\end{definition}

\begin{example}\label{aloha-eg}
As an example scenario that can be modelled as a CSG, consider a number of users trying to send packets using the slotted ALOHA protocol studied in~\cite{KNPS20b,KNPS21,KNPS22}. If there is a collision or if sending a packet fails, 
a user waits for some number of slots before resending, with the wait set according to an exponential backoff scheme. 


If we model this scenario as a CSG then, when a player has a packet to send, the actions available to the player correspond to either sending their packet or waiting to send the packet at some future time step. In the case when one coalition of players has an objective related to sending their packets efficiently, e.g., minimising the expected time to send their packets, and the remaining players form a second coalition and have the dual objective, we can model this scenario as a zero-sum CSG. In such a zero-sum CSG, the optimal strategy for the first coalition is to try and choose times to send that avoid collisions, while the second coalition will do the opposite and instead try and cause collisions. On the other hand, when there are more coalitions and each coalition's goal corresponds to sending their own packets efficiently, we can model this as a nonzero-sum CSG. Here we would be looking for equilibria, i.e., profiles such that no coalition could improve its objective by changing its strategy, which are also optimal, e.g., the sum of the expected times is minimal or the difference between the expected time to send for each coalition is minimal.
\end{example}
\section{Property specifications and model checking for CSGs}\label{logic-sect}


Probabilistic model checking is a technique for systematically constructing a stochastic model
and analysing it against a quantitative property formally specified in temporal logic.
This approach can be used either to \emph{verify} that a specification is always satisfied
or to perform \emph{strategy synthesis}, i.e., to construct a witness to the satisfaction of a property.
In the context of CSGs, the latter means synthesising strategies for one or more players (or coalitions)
such that the resulting behaviour of the game satisfies the specification.

To specify properties of labelled CSGs, we use the property specification language of the PRISM-games model checker~\cite{KNPS20},
which is based on the logic PCTL (probabilistic computation tree logic)~\cite{HJ94},
extended with operators to specify expected reward properties~\cite{FKNP11}
and the \emph{coalition} operator $\coalition{C}$ from alternating temporal logic (ATL)~\cite{alur2002alternating}.
The variant of this logic that just considers \emph{zero-sum} formulae
is referred to as rPATL (probabilistic alternating-time temporal logic with rewards) in~\cite{CFK+13b},
but here we use a further extended version that also supports
\emph{nonzero-sum} properties, using the notion of equilibria~\cite{KNPS21,KNPS22}.


\begin{definition}[PRISM-games logic \cite{KNPS21,KNPS22}]\label{def:logic}
The syntax of the PRISM-games logic is given by the grammar:
\begin{eqnarray*}
\phi & \coloneqq & \mathtt{true} \; \mid \; \ap \; \mid \; \neg \phi \; \mid \; \phi \wedge \phi \; \mid \; \coalition{C}\probop{\sim q}{\psi} \; \mid \; \coalition{C}\rewop{r}{\sim x}{\rho} \; \mid \;    \nashop{\mathbb{C}}{\star_1,\star_2}{\opt \sim x}{\theta} \\
\psi & \coloneqq & \next \phi \ \mid \ \phi \buntil \phi \ \mid \ \phi \until \phi \\
\rho & \coloneqq & \sinstant{=k} \ \mid \ \scumul{\leq k} 
\ \mid \  \future \phi \\
\theta & \coloneqq & \probop{}{\psi}{+}{\cdots}{+}\probop{}{\psi} \ \mid \  \rewop{r}{}{\rho}{+}{\cdots}{+}\rewop{r}{}{\rho}  
\end{eqnarray*}
where $\ap$ is an atomic proposition, $\mathbb{C} = C_1{:}\cdots{:}C_m$, $C$ and $C_1,\dots,C_m$ are coalitions of players such that $C'  =  N {\setminus} C$, $C_i \cap C_j = \emptyset$ for all $1\leq i \neq j \leq m$, $(\star_1,\star_2) \in \{\NE,\CE \} {\times} \{\SW, \SF \}$, $\opt \in \{ \min,\max\}$, $\sim \,\in \{<, \leq, \geq, >\}$, $q \in\Qset\cap[0, 1]$, $x \in \Qset$, $r$ is a reward structure and $k \in \Nset$. %
\end{definition}
The syntax distinguishes between state ($\phi$), path ($\psi$) and reward ($\rho$) formulae.
State formulae are evaluated over states of a CSG, while path and reward formulae are both evaluated over paths.
Sums of formulae ($\theta$) are used to specify multiple objectives for equilibria.

We omit the formal semantics, which can be found in \cite{KNPS21,KNPS22}.
Path and reward formulae are used to express the utilities of the players, i.e., random variables over paths. For path formulae, we allow \emph{next} ($\next \phi$), \emph{bounded until} ($\phi \buntil \phi$) and \emph{unbounded until} ($\phi \until \phi$).
We also allow the usual equivalences such as $\future\phi \equiv \true\until\phi$
(i.e., \emph{probabilistic reachability}) and $\bfuturep{k}\phi \equiv \true\buntil\phi$
(i.e., \emph{bounded probabilistic reachability}). The random variable corresponding to the path formula $\psi$ returns 1 for paths that satisfy $\psi$ and zero otherwise. For reward formulae, we allow instantaneous (state) reward at the $k$th step (\emph{instantaneous reward} $\sinstant{=k}$),
reward accumulated over $k$ steps (\emph{bounded cumulative reward} $\scumul{\leq k}$),
and reward accumulated until a formula $\phi$ is satisfied (\emph{expected reachability} $\future \phi$). The random variable corresponding to the reward formula $\rho$ returns for a path the reward corresponding to $\rho$.

\subsection{Zero-sum formulae}

A state satisfies a formula $\coalition{C} \probop{\sim q}{\psi}$ if the coalition of players $C\subseteq N$ can ensure that the probability of the path formula $\psi$ being satisfied is ${\sim} q$, regardless of the actions of the other players ($N{\setminus}C$) in the game.
A state satisfies a formula $\coalition{C} \rewop{r}{\sim x}{\rho}$ if the players in $C$ can ensure that the expected value of the reward formula $\rho$ for reward structure $r$ is ${\sim} x$, whatever the other players do.

The model checking algorithms presented in~\cite{KNPS21} involve graph-based analysis followed by backward induction~\cite{SW+01,NMK+44} for exact computation of \emph{finite-horizon properties} and value iteration~\cite{RF91,CH08} for approximate computation of \emph{infinite-horizon properties}. During both backward induction and value iteration for each state, at each iteration, an LP problem of size $|A|$ must be solved (corresponding to finding the value of a zero-sum one-shot game), which has complexity PTIME~\cite{Kar84}. 

Strategy synthesis for the formulae $\coalition{C} \probop{\sim q}{\psi}$ and $\coalition{C} \rewop{r}{\sim x}{\rho}$ corresponds to finding optimal strategies for the players in coalition $C$ when their objective, respectively, is maximising the probability of satisfying the formula $\psi$ and maximising the expected value of the reward formula with respect to the reward structure $r$. All strategies synthesised are randomised and can be found during model checking by extracting not just the value of the zero-sum one-shot game solved in each state, but also an optimal (randomised) strategy. For infinite-horizon objectives, the synthesised strategies are memoryless, while for finite-horizon objectives, the synthesised strategies are finite-memory, with a separate distribution required for each state and each time step.


\subsection{Nonzero-sum formulae}

Nonzero-sum formulae allow us to reason about equilibria,
for either of the types (NE or CE) and optimality criteria (SW or SF) considered here.
A probabilistic formula $\nashop{C_1{:}{\cdots}{:}C_m}{\star_1,\star_2}{\max \sim x}{\probop{}{\psi_1}{+}{\cdots}{+}\probop{}{\psi_m}}$ is true in a state if, when the players form the coalitions $C_1,\dots,C_m$,
there is a subgame-perfect equilibrium of type $\star_1$ meeting the optimality criterion $\star_2$  for which the \emph{sum} of the values of the objectives
$\probop{}{\psi_1},\dots,\probop{}{\psi_m}$ for the coalitions $C_1,\dots,C_m$
satisfies ${\sim} x$. The objective of coalition $C_i$ is to maximise the probability of satisfying
a path formula $\psi_i$.

For a reward formula $\nashop{C_1{:}{\cdots}{:}C_m}{\star_1,\star_2}{\max \sim x}{\rewop{r_1}{}{\rho_1}{+}{\cdots}{+}\rewop{r_m}{}{\rho_m}}$ the meaning is similar; however, here the objective of coalition $C_i$ refers to a reward formula $\rho_i$ with respect to reward structure $r_i$.
%
Formulae of the form $\nashop{C_1{:}{\cdots}{:}C_m}{\star_1,\star_2}{\min \sim x}{\theta}$ correspond to the dual notion of cost equilibria, which are also supported.
We also allow \emph{numerical} queries of the form $\nashop{C_1{:}{\cdots}{:}C_m}{\star_1,\star_2}{\opt =?}{\theta}$, which return the sum of the subgame-perfect equilibrium's values of of type $\star_1$ meeting the optimality criterion $\star_2$.

Model checking algorithms, presented in~\cite{KNPS21,KNPS20b,KNPS22}, involve solving an $m$-player \emph{coalition game} $\game^\cC$, where $\cC = \{C_1,\dots,C_m\}$ and the choices of each player $i$ in $\game^\cC$ correspond to the choices of the players in coalition $C_i$ in $\game$.
If all the objectives in $\theta$ are finite-horizon, then \emph{backward induction}~\cite{SW+01,NMK+44} can be applied to compute (precise) optimal equilibria values. On the other hand, if all the objectives are infinite-horizon, \emph{value iteration}~\cite{CH08} can be used to approximate optimal equilibria values. When there is a combination of finite- and infinite-horizon objectives, the game under study is modified in a standard manner to make all objectives infinite-horizon. 

Both backward induction and value iteration over the CSG $\game^\cC$ work by iteratively computing new values for each state $s$ of $\game^\cC$.
The values for each state, in each iteration,
are found by computing optimal equilibria values,  with respect to the criterion $\star_2$ and equilibrium type $\star_1$, of an NFG $\nfgame$ 
whose utility function is derived from the outgoing transition probabilities from $s$ in the CSG and the values computed for successor states of $s$ in the previous iteration.
%

We can synthesise a strategy profile representing the appropriate type of equilibrium for the CSG
by combining the optimal strategies for the equilibria generated in each individual state during solution.
As for zero-sum formulae, randomisation is required and memory is needed both to keep track of both the step bound of finite-horizon objectives and the satisfaction of each player's objective.

\begin{example}\label{aloha2-eg}
We now return to the scenario from \egref{aloha-eg}, where a number users are attempting to send packets using the slotted ALOHA protocol. The zero-sum formulae
$\coalition{\mathit{usr}_1,\dots,\mathit{usr}_k} \rewop{\mathit{time}}{\min=?}{\future \mathsf{sent}_{1\dots k}}$ and  $\coalition{\mathit{usr}_1,\dots,\mathit{usr}_k} \probop{\max=?}{\future \mathsf{sent}_{1\dots k} \wedge t \leq D}$
represent the case where the first $k$ users form a coalition and try to minimise the expected time to send their packets or maximise the probability they send their packets within a deadline $D \in \mathbb{N}$, respectively, while the remaining users form a second coalition and try and achieve the opposite objective, i.e., maximise the expected time or minimise the probability.

On the other hand, in the nonzero-sum case, if we suppose there are $m$ users and the objective of each user is to minimise the expected time to send their packet, this can be expressed by the nonzero-sum formula $\nashop{\mathit{usr}_1{:}\cdots{:}\mathit{usr}_m}{\star_1,\star_2}{\min=?}{\rewop{\mathit{time}}{}{\future \mathsf{sent}_1}{+}{\cdots}{+}\rewop{\mathit{time}}{}{\future \mathsf{sent}_m}}$.
\end{example}
\section{Tool support and case studies}

Tool support for the modelling and automated verification of CSGs has been implemented in PRISM-games~\cite{KNPS20}, which is available from~\cite{pgwww}. 
%
A variety of case studies have been modelled and analysed as CSGs with the tool, using both zero-sum and nonzero-sum properties. These 
include: a robot coordination problem~\cite{KNPS21}; futures market investors~\cite{KNPS21,KNPS22}; medium access control~\cite{KNPS20b,KNPS21}; power control~\cite{KNPS21,KNPS22}; a public good game~\cite{KNPS20b,KNPS22} and secret sharing~\cite{KNPS20b}. The results for these case studies demonstrate: the advantages of using CSGs for modelling (for example, with respect to simpler turn-based games);
that using nonzero-sum properties can yield gains
for the players (or coalitions); and that the use of correlated equilibria and social fairness results may be advantageous compared to Nash equilibria and social welfare.
We give a brief description of the functionality and implementation of PRISM-games
and then present a representative case study: the slotted ALOHA protocol.

\subsection{PRISM-games}

PRISM-games~\cite{KNPS20} is an extension of the PRISM model checker, which provides support
for a variety of stochastic game models, including turn-based and concurrent
multi-player stochastic games, and (turn-based) timed probabilistic games.

These are all described in the PRISM-games modelling language,
a stochastic extension of the Reactive Modules formalism~\cite{AH99}.
The language facilitates the specification of systems comprising multiple components, referred to as modules, that operate in parallel, both asynchronously and synchronously through action labels.
Each module has a number of finite-valued variables and a state of the system specifies the values of the variables of all modules. The behaviour of each module is defined by probabilistic guarded commands, where the guard is a predicate over the variables of the modules and the command specifies a probabilistic update of the module's variables.
In a CSG model, each player constitutes a set of modules,
and these therefore execute concurrently.

PRISM-games provides a graphical user interface for designing and simulating stochastic games models, but its core functionality is to exhaustively construct a game and perform verification and strategy synthesis against a logical specification.
For CSGs, the PRISM-games logic described in \defref{def:logic} is supported,
and the resulting strategies can be exported, simulated or further verified.

The implementation of CSG model checking is built within PRISM's `explicit' engine,
which is based on sparse matrices and implemented in Java.
Computing values (and optimal strategies) for zero-sum NFGs, needed for zero-sum formulae,
is performed using the LPSolve library~\cite{LPS} via linear programming.
The computation of SWNE or SFNE for nonzero-sum NFG, required for nonzero-sum formulae, depends on the number of players. For two players~\cite{KNPS21}, labelled polytopes are used to characterise and find NE values through a reduction to SMT in both Z3~\cite{Z3} and Yices~\cite{Dut14}.
If there are more than two players, the implementation~\cite{KNPS20b} is based on support enumeration and uses a combination of the SMT solver Z3~\cite{Z3} and the nonlinear optimisation suite {\sc Ipopt}~\cite{Wac09}. In the case of SWCE for nonzero-sum NFGs, as the problem reduces to an LP problem~\cite{KNPS22}, either Gurobi~\cite{gurobi} or the SMT solver Z3~\cite{Z3} is used. Finally, for SFCE, since the problem does not reduce directly to an LP problem, only Z3 can be used.

\subsection{The ALOHA case study}



We now return to the slotted ALOHA protocol discussed in Examples~\ref{aloha-eg} and \ref{aloha2-eg} to illustrate the benefits of game-theoretic analysis with CSGs. For further details of this case study, as well as several others, see \cite{pgwww}. Recall that, in the slotted ALOHA protocol, a number of users are attempting to send packets on a shared medium. 
We assume that, in any time slot, if a single user tries to send a packet then there is a probability ($q$) that the packet is sent and, as more users try and send, then the probability of success decreases. If sending a packet fails, the user waits for a number of slots before resending, defined according to an exponential backoff scheme. More precisely, each user maintains a backoff counter,  which it increases each time there is a failure (up to $b_{\max}$) and, if the counter equals $k$, randomly chooses the slots to wait from $\{0,1,\dots,2^k{-}1\}$. 

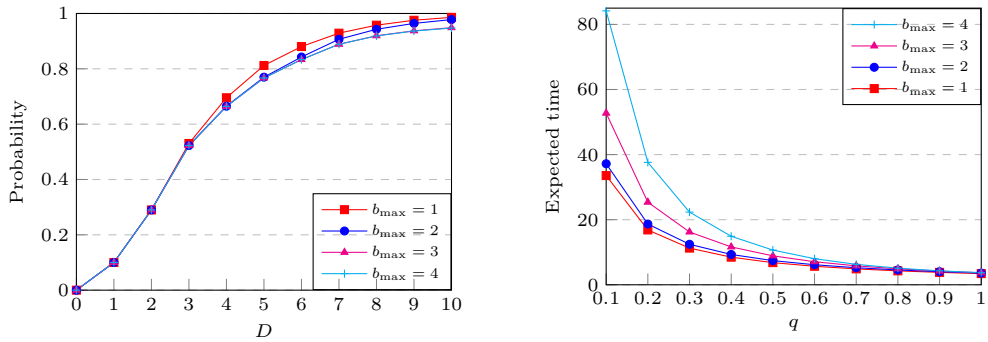
\begin{figure}[t]
\centering
\begin{subfigure}{.49\textwidth}
\centering
\scriptsize{

\begin{tikzpicture}
\begin{axis}[
    ylabel={Probability},
    xlabel={$D$},
    xmin=0, xmax=10,
    xtick={0,1,2,3,4,5,6,7,8,9,10},
    ymin=0, ymax=1,
    ytick={0,0.2,0.4,0.6,0.8,1.0},
    xtick pos=left,
    ytick pos=left,
    ymajorgrids=true,
    grid style=dashed,
    height=5.25cm,
    width=0.95\textwidth,
    legend entries={
                $b_{\max}=1$,
                $b_{\max}=2$,  
                $b_{\max}=3$,  
                $b_{\max}=4$  
                },
    legend style={fill=none, 
                    at={(1,0)}, anchor=south east, 
                    nodes={scale=0.8, transform shape}},
]
\addlegendimage{mark=square*,red,mark size=1.5pt}
\addlegendimage{mark=*,blue,mark size=1.5pt}
\addlegendimage{mark=triangle*,magenta,mark size=1.5pt}
\addlegendimage{mark=+,cyan,mark size=1.5pt}
]
\addplot[mark=square*,red,mark size=1.5pt] table [x=tmax, y=b1, col sep=comma]{csv/aloha/alh_bckff_zero_deadline.txt};
\addplot[mark=*,blue,mark size=1.5pt] table [x=tmax, y=b2, col sep=comma]{csv/aloha/alh_bckff_zero_deadline.txt};;

\addplot[mark=triangle*,magenta,mark size=1.5pt] table [x=tmax, y=b3, col sep=comma]{csv/aloha/alh_bckff_zero_deadline.txt};
\addplot[mark=+,cyan,mark size=1.5pt] table [x=tmax, y=b4, col sep=comma]{csv/aloha/alh_bckff_zero_deadline.txt};


\end{axis}
\end{tikzpicture}
}
\end{subfigure}
\begin{subfigure}{.49\textwidth}
\centering
\scriptsize{
\begin{tikzpicture}
\begin{axis}[
    ylabel={Expected time},
    xlabel={$q$},
    xmin=0.1, xmax=1,
    xtick={0.1,0.2,0.3,0.4,0.5,0.6,0.7,0.8,0.9,1.0},
    ymin=0, ymax=85,
    xtick pos=left,
    ytick pos=left,
    ymajorgrids=true,
    grid style=dashed,
    height=5.25cm,
    width=0.95\textwidth,
    legend entries={
                 $b_{\max}=4$,
                $b_{\max}=3$,  
                $b_{\max}=2$,  
                $b_{\max}=1$  
                },
    legend style={fill=none, 
                at={(1,1)}, anchor=north east, 
                nodes={scale=0.8, transform shape}}
]
\addlegendimage{mark=+,cyan,mark size=1.5pt}
\addlegendimage{mark=triangle*,magenta,mark size=1.5pt}
\addlegendimage{mark=*,blue,mark size=1.5pt}
\addlegendimage{mark=square*,red,mark size=1.5pt}
]

\addplot[mark=square*,red,mark size=1.5pt] table [x=q, y=b1, col sep=comma]{csv/aloha/alh_bckff_zero_expected.txt};
\addplot[mark=*,blue,mark size=1.5pt] table [x=q, y=b2, col sep=comma]{csv/aloha/alh_bckff_zero_expected.txt};;

\addplot[mark=triangle*,magenta,mark size=1.5pt] table [x=q, y=b3, col sep=comma]{csv/aloha/alh_bckff_zero_expected.txt};
\addplot[mark=+,cyan,mark size=1.5pt] table [x=q, y=b4, col sep=comma]{csv/aloha/alh_bckff_zero_expected.txt};

\end{axis}
\end{tikzpicture}
}
\end{subfigure}
\vspace*{-0.2cm}
\caption{Results from a CSG model of the ALOHA protocol:
one user maximising the probability of sending their packet before a deadline $D$ (left);
and minimising the expected time to send the packet, assuming a message transmission failure probability $q$ (right).}
\label{aloha-zero-fig}
\vspace*{-0.2cm}
\end{figure}

\subparagraph*{Zero-sum properties.}
We first consider the zero-sum properties $\coalition{\mathit{usr}_1} \probop{\max=?}{\futureop^{\leq D} \mathsf{sent}_1}$ and $\coalition{\mathit{usr}_1} \rewop{\mathit{time}}{\min=?}{\future \mathsf{sent}_1}$ from \egref{aloha2-eg}, which correspond to the first user trying to maximise the probability that their packet is sent before a deadline and trying to minimise the expected time to send their packet, respectively. The results for the first property when $q=0.9$ as the deadline $D$ varies, and for the second property as the probability $q$ varies, are presented in \figref{aloha-zero-fig} for different values of $b_{\max}$. We see that the probability decreases and the expected time decreases as $b_{\max}$ increases; this is because, as $b_{\max}$ increases, the additional time the first user can spend in backoff outweighs the gains in reducing the chance of avoiding further collisions. By performing strategy synthesis we see that it is optimal for the first user to initially randomly decide as to when to send their packet in order to avoid collisions with the coalition of the second and third user. However, this changes to a deterministic strategy of just sending its packet when the other users have sent their packets or the deadline is getting close, and therefore waiting will mean the deadline is missed. 


\begin{figure}[t]
\centering
\begin{subfigure}{.49\textwidth}
\centering
\scriptsize{
\begin{tikzpicture}
\begin{axis}[
    ylabel={Sum of Probabilities},
    xlabel={$D$},
    xmin=0, xmax=10,
    xtick={0,1,2,3,4,5,6,7,8,9,10},
    ymin=0, ymax=2,
    ytick={0,0.5,1,1.5,2.0,2.5},
    xtick pos=left,
    ytick pos=left,
    ymajorgrids=true,
    grid style=dashed,
    height=5.25cm,
    width=0.95\textwidth,
    legend entries={
                {$\coalition{\mathit{usr}_{\scale{.75}{1}}{:}\mathit{usr}_{\scale{.75}{2}}}_{=?}(\mathtt{P}{+}\mathtt{P})$},
                \textit{$\coalition{\mathit{\mathit{usr}_{\scale{.75}{1}}}}{\mathtt P_{\scale{.75}{\max}}}+{\mathtt P_{\scale{.75}{\max}}}$},
                \textit{$\coalition{\mathit{\mathit{usr}_{\scale{.75}{1}}}}{\mathtt P_{\scale{.75}{\max}}}+{\mathtt P_{\scale{.75}{\min}}}$}    
                },
    legend style={fill=none, 
                    at={(1,0)}, anchor=south east, 
                    nodes={scale=0.8, transform shape}},
]
\addlegendimage{mark=square*,red,mark size=1.5pt}
\addlegendimage{mark=*,blue,mark size=1.5pt}
\addlegendimage{mark=triangle,orange,mark size=1.5pt}
]

\addplot[mark=square*,red,mark size=1.5pt] table [x=tmax, y=peq, col sep=comma]{csv/aloha/alh_bckff_3p_tmax.csv};
\addplot[mark=*,blue,mark size=1.5pt] table [x=tmax, y=pmxmx, col sep=comma]{csv/aloha/alh_bckff_3p_tmax.csv};
\addplot[mark=triangle,orange,mark size=1.5pt] table [x=tmax, y=pmxmn, col sep=comma]{csv/aloha/alh_bckff_3p_tmax.csv};


\end{axis}
\end{tikzpicture}
}
\end{subfigure}
\begin{subfigure}{.49\textwidth}
\centering
\scriptsize{
\begin{tikzpicture}
\begin{axis}[
    ylabel={Probability},
    xlabel={$q$},
    xmin=0, xmax=1,
    xtick={0,0.1,0.2,0.3,0.4,0.5,0.6,0.7,0.8,0.9,1.0},
    ymin=0, ymax=1,
    ytick={0,0.2,0.4,0.6,0.8,1.0},
    xtick pos=left,
    ytick pos=left,
    ymajorgrids=true,
    grid style=dashed,
    height=5.25cm,
    width=0.95\textwidth,
    legend entries={
                $\mathit{usr}_{\scale{.75}{1}}$ (SWNE),
                $\mathit{usr}_{\scale{.75}{2}}$ (SWNE),
                $\mathit{usr}_{\scale{.75}{1}}$ (Max),
                $\mathit{usr}_{\scale{.75}{2}}$ (Max)
                },
    legend style={fill=none, 
                at={(0,1)}, anchor=north west, 
                nodes={scale=0.8, transform shape}}
]
\addlegendimage{mark=square*,red,mark size=1.5pt}
\addlegendimage{mark=*,blue,mark size=1.5pt}
\addlegendimage{mark=triangle*,magenta,mark size=1.5pt}
\addlegendimage{mark=+,cyan,mark size=1.5pt}
]

\addplot[mark=square*,red,mark size=1.5pt] table [x=p11, y=player1, col sep=comma]{csv/aloha/alh_bckff_3p_p11_players.csv};
\addplot[mark=*,blue,mark size=1.5pt] table [x=p11, y=player2, col sep=comma]{csv/aloha/alh_bckff_3p_p11_players.csv};;

\addplot[mark=triangle*,magenta,mark size=1.5pt] table [x=p11, y=p1, col sep=comma]{csv/aloha/alh_bckff_3p_p11_players.csv};
\addplot[mark=+,cyan,mark size=1.5pt] table [x=p11, y=p2, col sep=comma]{csv/aloha/alh_bckff_3p_p11_players.csv};

\end{axis}
\end{tikzpicture}
}
\end{subfigure}
\vspace*{-0.2cm}
\caption{Results from CSG equilibria synthesis on the ALOHA protocol,
maximising the probabilities of sending packets by deadline $D$
for two coalitions (user 1, and users 2 and 3):
probability sums (left) and individual probabilities (right).
}
\label{aloha-deadline-fig}
\vspace*{-0.2cm}
\end{figure}
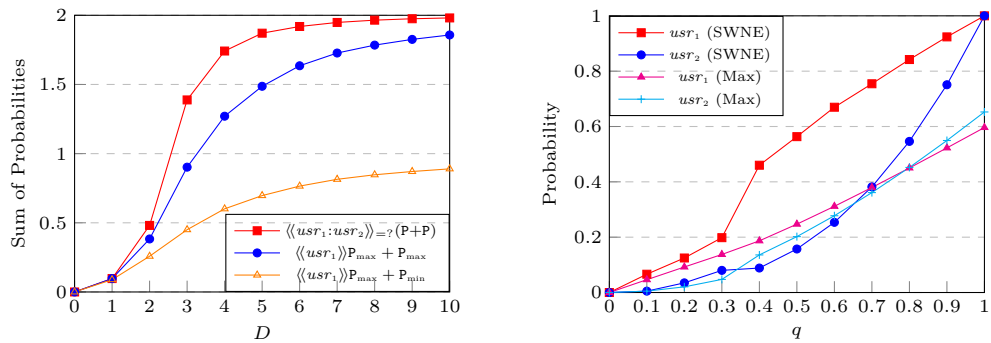

\subparagraph*{Benefits of equilibria.}
We next highlight the analysis from \cite{KNPS21}, which demonstrates the advantages of cooperation through nonzero-sum properties when using Nash equilibria (NE) and the social welfare (SW) optimality criterion, as opposed to adopting a strategy that assumes antagonistic behaviour. The first non-zero sum property we consider corresponds to the case when each user is trying to maximise the probability of sending their packet before a deadline $D$, with users 2 and 3 forming a coalition, represented by the formula $\nashop{\mathit{usr}_1{:}\mathit{usr}_2{,}\mathit{usr}_3}{\NE,\SW}{\max=?}{\probop{}{\future (\mathsf{sent}_1 \wedge t {\leq} D)}+\probop{}{\future (\mathsf{sent}_2 \wedge \mathsf{sent}_3 \wedge t {\leq} D)}}$.

\figref{aloha-deadline-fig} presents total values (the sum of the probabilities for user 1 and the coalition of user 2 and 3) as $D$ varies (left) and individual values as $q$ varies (right). By performing strategy synthesis, the analysis found that the collaboration is dependent on both $D$ and $q$. In particular, if the users have more  time there is a greater chance for the users to collaborate by sending in different slots, 
whereas, when $q$ is large, it is unlikely users need to repeatedly send, so again can send in different slots. 
As \figref{aloha-deadline-fig} (right) demonstrates, since the coalition has more packets to send, their probabilities are lower.

\subparagraph*{Equilibria types and optimality criteria.}
Finally, we report on the experiments of \cite{KNPS22},
which investigate the benefits of using different types of equilibria,
i.e., \emph{correlated} (CE) over \emph{Nash} equilibria,
and optimality criteria, i.e., \emph{social fairness} (SF) over \emph{social welfare} (SW).
The experiments varied the number of users and considered the case when the objective of each individual user is to minimise the expected time to send their packet, which is represented by the nonzero-sum formula $\nashop{\mathit{usr}_1{:}\cdots{:}\mathit{usr}_m}{\star_1,\star_2}{\min=?}{\rewop{\mathit{time}}{}{\future \mathsf{sent}_1}{+}{\cdots}{+}\rewop{\mathit{time}}{}{\future \mathsf{sent}_m}}$.

Synthesising optimal strategies for this specification, it was found that the cases for SWNE and SWCE coincide (although 
SWCE 
returns a joint strategy for the users, this joint strategy can be separated to form a strategy profile). This profile required one user to try and send first,  and then for the remaining users to take turns to try and send afterwards. If a user fails to send, then they enter backoff and allow all remaining users to try and send before trying to send again. The reason for this is that there is no gain in a user trying to send at the same time as another user, as this will increase the probability of a collision and thus their packets not being sent, and therefore the users having to spend time in backoff.

For SFNE, which has only been implemented for the two-player case, the two users followed identical strategies, which involve randomly deciding whether to wait or transmit, unless they are the only user that has not transmitted, and then they always try to send when not in backoff. In the case of SFCE,  users employed a shared probabilistic signal to coordinate which user sends next. Initially, this was a uniform choice over the users, but as time progresses the signal favoured the users with lower backoff counters as these users had fewer opportunities to send their packet previously. 

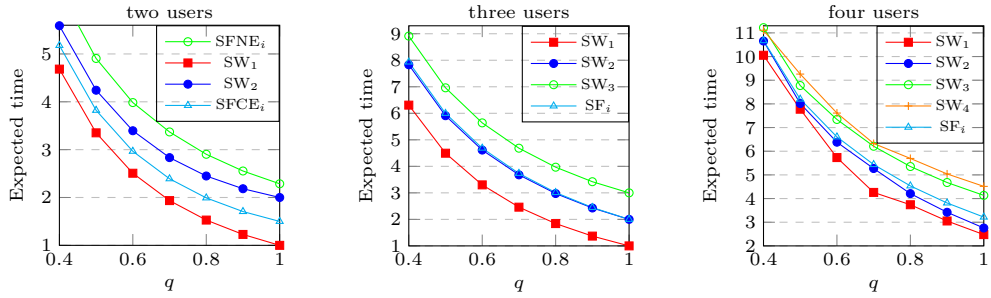
\begin{figure}[t]
\begin{subfigure}{.32\textwidth}
\centering
\scriptsize{
\begin{tikzpicture}
\begin{axis}[
    title style={yshift=-2ex},
    title={two users},
    ylabel={Expected time},
    xlabel={$q$},
    xmin=0.4, xmax=1,
    xtick={0.4,0.6,0.8,1.0},
    ymin=1, ymax=5.6,
    ytick={1,2,3,4,5},
    legend pos=north east,
    legend style={fill=none},
    ymajorgrids=true,
    grid style=dashed,
    height=4.5cm,
    width=\textwidth,
    legend entries={
                $\text{SFNE}_i$,
                $\text{SW}_1$,
                $\text{SW}_2$,
                $\text{SFCE}_i$,
    },
    legend style={fill = none, 
            at={(1,1)}, anchor=north east, 
            nodes={scale=.8, transform shape}},
]
\addlegendimage{mark=o,green,mark size=1.5pt}
\addlegendimage{mark=square*,red,mark size=1.5pt}
\addlegendimage{mark=*,blue,mark size=1.5pt}
\addlegendimage{mark=triangle,cyan,mark size=1.5pt}
]

\addplot[mark=o,green,mark size=1.5pt] table [x=q, y=nf, col sep=comma]{csv/aloha2_expected.txt};
\addplot[mark=square*,red,mark size=1.5pt] table [x=q, y=sw1, col sep=comma]{csv/aloha2_expected.txt};
\addplot[mark=*,blue,mark size=1.5pt] table [x=q, y=sw2, col sep=comma]{csv/aloha2_expected.txt};
\addplot[mark=triangle,cyan,mark size=1.5pt] table [x=q, y=f, col sep=comma]{csv/aloha2_expected.txt};

\end{axis}
\end{tikzpicture}
}
\end{subfigure}
\begin{subfigure}{.32\textwidth}
\centering
\scriptsize{
\begin{tikzpicture}
\begin{axis}[
    title style={yshift=-2ex},
    title={three users},
    ylabel={Expected time},
    xlabel={$q$},
    xmin=0.4, xmax=1,
    xtick={0.4,0.6,0.8,1.0},
    ymin=1, ymax=9.3,
    ytick={1,2,3,4,5,6,7,8,9},
    legend pos=north east,
    legend style={fill=none},
    ymajorgrids=true,
    grid style=dashed,
    height=4.5cm,
    width=\textwidth,
    legend entries={
                $\text{SW}_1$,
                $\text{SW}_2$,
                $\text{SW}_3$,
                $\text{SF}_i$,
    },
    legend style={fill = none, 
            at={(1,1)}, anchor=north east, 
            nodes={scale=.8, transform shape}},
]
\addlegendimage{mark=square*,red,mark size=1.5pt}
\addlegendimage{mark=*,blue,mark size=1.5pt}
\addlegendimage{mark=o,green,mark size=1.5pt}
\addlegendimage{mark=triangle,cyan,mark size=1.5pt}
]

\addplot[mark=square*,red,mark size=1.5pt] table [x=q, y=sw1, col sep=comma]{csv/aloha3_expected.txt};
\addplot[mark=*,blue,mark size=1.5pt] table [x=q, y=sw2, col sep=comma]{csv/aloha3_expected.txt};
\addplot[mark=o,green,mark size=1.5pt] table [x=q, y=sw3, col sep=comma]{csv/aloha3_expected.txt};
\addplot[mark=triangle,cyan,mark size=1.5pt] table [x=q, y=f, col sep=comma]{csv/aloha3_expected.txt};

\end{axis}
\end{tikzpicture}
}
\end{subfigure}
\begin{subfigure}{.32\textwidth}
\centering
\scriptsize{
\begin{tikzpicture}
\begin{axis}[
    title style={yshift=-2ex},
    title={four users},
    ylabel={Expected time},
    xlabel={$q$},
    xmin=0.4, xmax=1,
    xtick={0.4,0.6,0.8,1.0},
    ymin=2, ymax=11.3,
    ytick={1,2,3,4,5,6,7,8,9,10,11},
    legend pos=north east,
    legend style={fill=none},
    ymajorgrids=true,
    grid style=dashed,
    height=4.5cm,
    width=\textwidth,
    legend entries={
                $\text{SW}_1$,
                $\text{SW}_2$,
                $\text{SW}_3$,
                $\text{SW}_4$,
                $\text{SF}_i$,
    },
    legend style={fill = none, 
            at={(1,1)}, anchor=north east, 
            nodes={scale=.8, transform shape}},
]
\addlegendimage{mark=square*,red,mark size=1.5pt}
\addlegendimage{mark=*,blue,mark size=1.5pt}
\addlegendimage{mark=o,green,mark size=1.5pt}
\addlegendimage{mark=+,orange,mark size=1.5pt}
\addlegendimage{mark=triangle,cyan,mark size=1.5pt}
]

\addplot[mark=square*,red,mark size=1.5pt] table [x=q, y=sw1, col sep=comma]{csv/aloha4_expected.txt};
\addplot[mark=*,blue,mark size=1.5pt] table [x=q, y=sw2, col sep=comma]{csv/aloha4_expected.txt};
\addplot[mark=o,green,mark size=1.5pt] table [x=q, y=sw3, col sep=comma]{csv/aloha4_expected.txt};
\addplot[mark=+,orange,mark size=1.5pt] table [x=q, y=sw4, col sep=comma]{csv/aloha4_expected.txt};
\addplot[mark=triangle,cyan,mark size=1.5pt] table [x=q, y=f, col sep=comma]{csv/aloha4_expected.txt};

\end{axis}
\end{tikzpicture}
}
\end{subfigure}
\vspace*{-0.3cm}
\caption{Results from different types of equilibria (correlated vs. Nash)
and optimality criteria (social fairness vs. social welfare)
for minimising the expected times for users to send packets in the ALOHA protocol.,
for varying numbers of users.}
\label{aloha234-fig}
\vspace*{-0.2cm}
\end{figure}

\figref{aloha234-fig} plots the optimal values for the users, where $\text{SW}_i$ corresponds to the optimal values (expected times to send their packets) for user $i$ for both SWNE and SWCE 
for the cases of two, three and four users. We see that the optimal values for the different users under SFNE and SFCE coincide, while under SWNE and SWCE they are different for each user (with the user sending first having the lowest and the user sending last the highest). Comparing the sum of the SWNE (and SWCE) values and that of the SFCE values, we see a small decrease in the sum of less than 2\% of the total, whereas for SFNE there is a greater difference as the users cannot coordinate, and hence try and send at the same time.

\section{Recent Developments: Neuro-symbolic CSGs}

The recent encouraging advances of AI, and particularly deep learning, have resulted in computing architectures that integrate components that are synthesized from data (e.g., implemented as neural networks)
with conventional, symbolic modules (e.g., controllers). Design automation support for such \emph{neuro-symbolic} systems is, however, lacking. To this end, we have developed the model of \emph{neuro-symbolic concurrent stochastic games} (NS-CSGs)~\cite{nscsgs,YSD+22},
which is targeted at AI-based autonomous systems, e.g., autonomous driving or aircraft controllers.
NS-CSGs are a variant of (continuous-space) CSGs, in which each player is a neuro-symbolic \emph{agent} and the agents act concurrently in a shared, continuous-state environment. As for the players of CSGs, each agent has a finite set of available actions and agents choose their actions simultaneously; however, in NS-CSGs the action choices cause the agents' local states to be updated probabilistically and the agents are endowed with a perception mechanism implemented as a neural network, through which they can observe the local states of the other agents and that of the environment and encode these observations as locally stored \emph{percepts}. The global states of NS-CSGs comprise the state of the environment together with the local state and percept of each agent, and are therefore infinite-state, in contrast to the CSGs discussed in the rest of this paper.


\begin{definition}[Neuro-symbolic concurrent stochastic game~\cite{nscsgs,YSD+22}]\label{defi:NS-CSG}
A \emph{neuro-symbolic concurrent stochastic multi-player game} (NS-CSG) $\nscsg$ comprises
players $(\agent_i)_{i \in N}$, for $N=\{1, \dots, n \}$, and an environment $E$ where:
\[
\agent_i  = (S_i,A_i,\Delta_i,\obs_i,\delta_i) \; \mbox{for $i \in N$}, \; \;
E = (S_E,\delta_E)
\]
and we have:
\begin{itemize}
\item $S_i = \Loc_i\times \Per_i$ is a set of states for $\agent_i$, where $\Loc_i \subseteq \mathbb{R}^{b_i}$ and $\Per_i \subseteq \mathbb{R}^{d_i}$ are finite sets of local states and percepts, respectively;
    
\item $S_E\subseteq \mathbb{R}^e$ is a closed infinite set of environment states;
  
\item $A_i$ is a nonempty finite set of actions for $\agent_i$ and $A \coloneqq (A_1 \cup \{ \bot \}) \times \cdots \times (A_n \cup \{ \bot \})$ is the set of \emph{joint actions}, where $\bot$ is an idle action disjoint from $\cup_{i=1}^n A_i$;

\item $\Delta_i: S_i \to 2^{A_i}$ is an available action function, defining the actions $\agent_i$ can take in each state;

\item $\obs_i : (\Loc_1 \times \cdots \times \Loc_n \times S_E)\to \Per_i$ is a perception function for $\agent_i$, mapping the local states of all agents and the environment to a percept of the agent, implemented via a neural network (NN) classifier; 
    
\item $\delta_i:S_i \times A \to \mathbb{P}(\Loc_i)$ is a partial probabilistic transition function for $\agent_i$ determining the distribution over the agent's local states given its current state and joint action;

\item $\delta_E:S_E \times A \to S_E$ is a partial deterministic environment transition function determining the environment's next state given its current state and joint action.
\end{itemize}
\end{definition}


A (global) state for an NS-CSG $\nscsg$ comprises a state $s_i = (\loc_i, \per_i)$ for each agent $\agent_i$ (a pair of a local state and percept) and an environment state $s_E$. If an NS-CSG is in a state $s=(s_1,\dots,s_n,s_E)$, then each $\agent_i$ simultaneously chooses one of the actions available in its state $s_i$ (if no action is available, i.e., $\Delta_i(s_i)=\emptyset$, it picks the idle action~$\bot$) yielding a joint action $\alpha=(a_1,\dots,a_n)\in A$. Next, each $\agent_i$ updates its local state to $\loc_i'\in \Loc_i$,
according to its probabilistic local transition function $\delta_i$, applied to its current state $(\loc_i,\per_i)$ and the joint action $\alpha$. The environment updates its state to $s_E'\in S_E$ according to its local deterministic transition function $\delta_E$, based on its state $s_E$ and on $\alpha$.  Finally, each agent, based on its new local state $\loc_i'$, observes the new local states of the other agents and environment through its perception function $\obs_i$ to generate a new percept $\per_i' = \obs_i(\loc_1',\dots,\loc_n',s_E')$. Thus, the game reaches the state $s'=(s_1',\dots,s_n', s_E')$, where $s_i' = (\loc_i', \per_i')$ for $i \in N$.
We assume that each perception function $\obs_i$ is implemented via an NN $f_i:\mathbb{R}^{b+e} \to \mathbb{P}(\Per_i)$ yielding a normalised score over different percept values, where $b = \sum_{i =1}^n b_i$; however, it can be any function including other types of machine learning models. A rule is then applied that selects the percept value with the maximum score. 

Formally, the semantics of an NS-CSG $\nscsg$ is given by an infinite-state CSG
$\sem{\nscsg}$ over the product of the states of the agents and environment, which assumes a particular structure of the transition function that distinguishes between agent and environment states and
uses the NN-based perception function to define
which states have the same characteristics.

\begin{definition}[Semantics of an NS-CSG]\label{semantics-def}
Given an NS-CSG $\nscsg$ consisting of $n$ players and an environment, the semantics of $\nscsg$ is $\sem{\nscsg} = (N,S,A,\Delta,\delta)$ where:
\begin{itemize}
\item $S=S_1\times \cdots \times S_n \times S_E$ is the set of (global) states, which contain both discrete and continuous elements;
\item $A = (A_1 \cup \{ \bot \}) \times \cdots \times (A_n \cup \{ \bot \})$;
\item $\Delta(s_1,\dots,s_n,s_E) = \cup_{i=1}^n \Delta_i(s_i)$;
\item $\delta: (S \times ((A_1 \cup \{\bot \}) \times \cdots \times (A_n  \cup \{\bot \}))) \to \mathbb{P}(S)$ is the partial probabilistic transition function, where for states $s=(s_1,\dots,s_n,s_E),s'=(s_1',\dots,s_n',s_E')\in S$ and joint action $\alpha=(a_1,\dots,a_n)\in A$, if $a_i \in \Delta_i(s_i)$ when $\Delta_i(s_i) \neq \emptyset$ and $a_i= \bot$ otherwise, then $\delta(s,\alpha)$ is defined and if $s_i'=(\loc_i',\per_i')$, $\per_i'=\obs_i(\loc_1',\dots,\loc_n',s_E')$ for all $i \in N$  and $s_E'=\delta_E(s_E,\alpha)$,  then
\[
            \delta(s,\alpha)(s')= \left( \mbox{$\prod_{i=1}^n$} \delta_i(s_i,\alpha)(\loc'_i) \right)
\]
and otherwise $\delta(s,\alpha)(s')=0$.
\end{itemize}
\end{definition}


To illustrate NS-CSGs, we present the VerticalCAS Collision Avoidance Scenario (VCAS[2]) \cite{KDJ-MJK:19,KDJ-SS-JBJ-MJK:19}, modelled in~\cite{YSD+22}, which is a variant of the one studied in \cite{ABKL+20-2}, where the agents' trust level is modelled probabilistically to account for possible uncertainty.

\begin{example}\label{vcas-example}
 The geometry of the VCAS[2] case study is shown in Figure~\ref{fig:vcas-geometry}. There are
 two aircraft (ownship and intruder), constituting the agents of the NS-CSG, both equipped with an NN-controlled collision avoidance system called VCAS. At each time unit (i.e., every second), VCAS issues an advisory ($\mathit{ad} \in \{1,\dots,9\}$) from which, together with the current trust level ($\mathit{tr} \in \{1,\dots,4\}$) from the previous advisory, the pilot needs to make a decision about the rate of acceleration, aimed at avoiding a near mid-air collision (NMAC) \cite{ABKL+20}.
 

\usetikzlibrary{snakes}
\begin{figure}[t]
		\centering
		\begin{tikzpicture}
		\node [aircraft side,fill=blue,minimum width=1cm,rotate=15] at (0,0) {};
		\node [aircraft side,fill=red,minimum width=1cm,xscale=-1] at (4.7,1) {}; 
		
		\draw[dashed] (0.5,0) -- (4.5,0);
		\draw[dashed] (4.2,-0.2) -- (4.2,1.2);
		\draw[snake=brace,raise snake=3pt,mirror snake, gap around snake=0.4mm,segment aspect=0.5] (0.5,0) -- (4.2,0)node [midway,yshift=-10pt] {$t$};
		\draw[snake=brace,raise snake=3pt, gap around snake=0.4mm, mirror snake] (4.2,0) -- (4.2,1)node [midway,xshift=11pt] {$h$};
		
		\draw[-stealth] (-1.1,-0.3) to node[left]{$\dot{h}_{\mathit{own}}$} (-1.1,0.3);
		\draw[-stealth] (5.7,0.7) to node[right]{$\dot{h}_{\mathit{int}}$} (5.7,1.3);
		
		\node at (0,0.5) {$(\mathit{tr}_{\mathit{own}},\mathit{ad}_{\mathit{own}})$};
		\node at (4.7,1.5) {$(\mathit{tr}_{\mathit{int}},\mathit{ad}_{\mathit{int}})$};
		\end{tikzpicture}
		\caption{Geometry with trust levels and advisories for the agents of the VCAS[2] case study.}
		\label{fig:vcas-geometry}
\vspace*{-0.2cm}
	\end{figure}
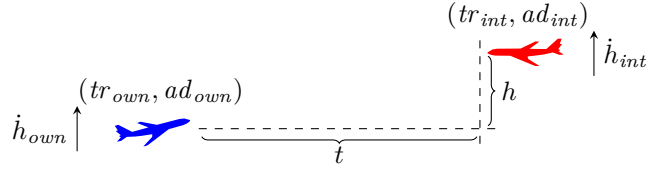

 The input to the VCAS system is the tuple $(h,\dot{h}_{\mathit{own}},\dot{h}_{\mathit{int}},t) \in \mathbb{R}^4$ including the relative altitude $h$ of the aircraft, the climb rate $\dot{h}_{\mathit{own}}$ of ownship, the climb rate $\dot{h}_{\mathit{int}}$ of intruder, and the time $t$ until loss of horizontal separation between the aircraft. VCAS is implemented via nine feed-forward NNs $F=\{ f_i:\mathbb{R}^4\to\mathbb{R}^9  \mid 1 \leq i \leq 9 \}$, each of which corresponds to an advisory and outputs the scores of the nine possible advisories. Each advisory will provide a set of accelerations for the agent to select from and the trust level increases probabilistically if the current advisory is compliant with the executed accelerations, and decreases otherwise. We formulate the NS-CSG with agents $\agent_i$ for $i\in \{\mathit{own}, \mathit{int} \}$ as follows:
\begin{itemize}
    \item the set of states for $\agent_i$ is given by $S_i = \{1,\dots,4\} \times \{1,\dots,9\}$,  where the agent state $s_i=(\mathit{tr}_i, \mathit{ad}_i) \in S_i$ has local state (trust level) $\mathit{tr}_i$ and percept (advisory) $\mathit{ad}_i$;
    
    \item the set of environment states is given by $S_E = \mathbb{R}^4$, where $s_E=(h,\dot{h}_{\mathit{own}},\dot{h}_{\mathit{int}},t) \in S_E$ represents the relative altitude, the climb rate of the ownship, the climb rate of the intruder, and the time until loss of their horizontal separation;
    
    \item the set of actions of $\agent_i$ is given by $A_i=\{0,\pm3.0, \pm 7.33, \pm 9.33, \pm 9.7, \pm11.7\}$ representing the acceleration options of $\agent_i$;
    
    \item the available action function $\Delta_i : \Per_i \rightarrow A_i$ of $\agent_i$ is independent of the local state of the agent and returns the set consisting two non-zero acceleration actions from Table~\ref{tab:advisory} for a given percept and the zero acceleration action; 
    
    
    \item  the perception function $obs_i : \Per_i \times S_E \rightarrow \{1,\dots,9\}$ of $\agent_i$ is independent of the local state of the agent and is given by the feed forward NNs $F$ of VCAS;
    
    
    \item the local transition function $\delta_i$ of $\agent_i$ updates the agent's trust level probabilistically according to its current trust level, its current advisory and its executed action;
    
    

    
    \item the environment transition function $\delta_E$ is given by $\delta_E((h,\dot{h}_{\mathit{own}},\dot{h}_{\mathit{int}},t),(\ddot{h}_\mathit{own},\ddot{h}_\mathit{int}))=(h',\dot{h}_{\mathit{own}}',\dot{h}_{\mathit{int}}',t')$ where for time step $\Delta t=1$: \begin{itemize}
    \item
    $h'=h-\Delta t(\dot{h}_{\mathit{own}}-\dot{h}_{\mathit{int}})-0.5\Delta t^2(\ddot{h}_{\mathit{own}}-\ddot{h}_{\mathit{int}})$;
    \item
    $\dot{h}_{\mathit{own}}'=\dot{h}_{\mathit{own}}+\ddot{h}_{\mathit{own}}\Delta t$;
    \item
    $\dot{h}_{\mathit{int}}'=\dot{h}_{\mathit{int}}+\ddot{h}_{\mathit{int}}\Delta t$;
    \item
    $t'=t-\Delta t$.
    \end{itemize}
\end{itemize}
\end{example}

\begin{table}[t]
\renewcommand{\arraystretch}{1.5}
\setlength{\tabcolsep}{2pt}  
\centering
\scriptsize{
\begin{tabular}{|c|l|l|c|c|} \hline
Label & \multirow{2}{*}{Advisory}  & \multirow{2}{*}{Description}  & \multirow{1}{*}{Vertical Range}   & \multirow{1}{*}{Actions} \\
$(ad_i)$ & & & \multirow{1}{*}{(Min, Max) ft/min} & \multirow{1}{*}{$\textup{ft/s}^2$}
 \\ \hline \hline	
1 &  COC  & Clear of Conflict & $(-\infty,+\infty)$& $-3$, $+3$
\\
2 & DNC  & Do Not Climb & $(-\infty,0]$& $-9.33$, $-7.33$
\\
3 &  DND & Do Not Descend & $[0,+\infty)$ & $7.33$, $+9.33$
\\
4 & DES1500 & Descend at least 1500 ft/min &  $(-\infty,-1500]$ &  $-9.33$, $-7.33$
\\
5 & CL1500 & Climb at least 1500 ft/min & $[+1500,+\infty)$ &  $+7.33$, $+9.33$
\\
6 & SDES1500 & Strengthen Descend to at least 1500 ft/min & $(-\infty,-1500]$ &  $-11.7$, $-9.7$
\\
7 & SCL1500 & Strengthen Climb to at least 1500 ft/min & $[+1500,+\infty)$ &  $+9.7$, $+11.7$
\\
8 & SDES2500 & Strengthen Descend to at least 2500 ft/min & $(-\infty,-2500]$ & $-11.7$, $-9.7$
\\ 
9 & SCL2500 & Strengthen Climb to at least 2500 ft/min & $[+2500,+\infty)$ & $+9.7$, $+11.7$
\\ \hline
\end{tabular}}
\vspace*{0.2cm}
\caption{Actions available for the agents of VCAS[2] for each advisory~\cite{ABKL+20-2}.}
\label{tab:advisory}
\vspace*{-0.4cm}
\end{table}

\subsection{Zero-sum NS-CSGs}
In view of the uncountable state spaces, \cite{nscsgs} presents an approach for zero-sum \emph{discounted} \emph{infinite-horizon} cumulative rewards under the assumption of full state observability for NS-CSGs, which exploits Borel state space decomposition and identifies model restrictions to ensure determinacy, and therefore existence of a value that corresponds to a unique fixed point. Value iteration and policy iteration algorithms to compute values and synthesise optimal strategies are also derived based on formulating piecewise linear or constant representations of the value functions and strategies for NS-CSGs.

\subsection{Nonzero-sum NS-CSGs}
In the case of nonzero-sum NS-CSGs, \cite{YSD+22} studies the \emph{undiscounted}, \emph{finite-horizon} equilibria synthesis problem. The use of finite-horizon objectives simplifies the analysis
(note that the existence of infinite-horizon NE for CSGs is an open problem~\cite{BMS14},
and the verification of non-probabilistic infinite-horizon reachability properties for neuro-symbolic games is undecidable~\cite{ABKL+20-2}). Both NE and CE using the SW optimality criteria are considered.
The algorithms, based on backward induction and non-linear programming, compute \emph{globally optimal} equilibria which,
from a fixed initial state, are optimal over the chosen time horizon, in contrast to the local optimality of equilibria for finite-state CSGs~\cite{KNPS21,KNPS22}.

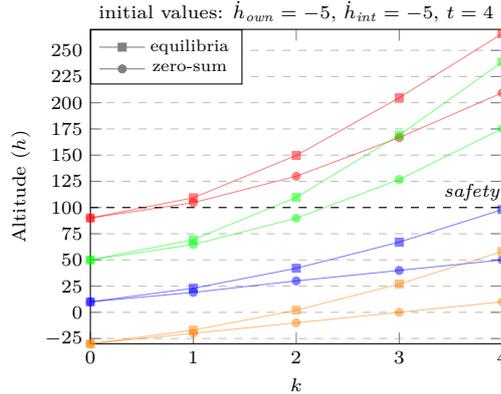
\begin{figure}[t]
\centering
\scriptsize{
\hspace{-0.0cm}
\trimbox{0.2cm 0.4cm 0.0cm 0.2cm}{ 
\begin{tikzpicture}
\begin{axis}[
    title style={yshift=-2ex},
    title={initial values: $\dot{h}_{\mathit{own}}=-5$, $\dot{h}_{\mathit{int}}=-5$, $t=4$},
    ylabel={Altitude ($h$)},
    xlabel={$k$},
    xmin=0, xmax=4.0,
    xtick={4,3,2,1,0},
    ymin=-30, ymax=270,
    ytick={-25,0,25,50,75,100,125,150,175,200,225,250},
    ymajorgrids=true,
    grid style=dashed,
    height=5.75cm,
    width=0.5\textwidth,
    legend entries={
                {equilibria},
                {zero-sum}
                },
    legend style={at={(0.0,1.0)},
                anchor=north west, 
                nodes={scale=0.9, transform shape}}            
]
\addlegendimage{mark=square*,gray,mark size=1.5pt}
\addlegendimage{mark=*,gray,mark size=1.5pt}
]

\draw [dashed] 
        (axis cs: 0,100) -- (axis cs: 4,100)
        node[pos=0.84, above right] {\emph{safety}};

\addplot[mark=square*,red,opacity=0.5,mark size=1.5pt] table [x=tau, y=pos_eq, col sep=comma]{csv/vcas_pos_90.txt};
\addplot[mark=*,red,opacity=0.5,mark size=1.5pt] table [x=tau, y=pos_zs, col sep=comma]{csv/vcas_pos_90.txt};

\addplot[mark=square*,green,opacity=0.5,mark size=1.5pt] table [x=tau, y=pos_eq, col sep=comma]{csv/vcas_pos_50.txt};
\addplot[mark=*,green,opacity=0.5,mark size=1.5pt] table [x=tau, y=pos_zs, col sep=comma]{csv/vcas_pos_50.txt};

\addplot[mark=square*,blue,opacity=0.5,mark size=1.5pt] table [x=tau, y=pos_eq, col sep=comma]{csv/vcas_pos_10.txt};
\addplot[mark=*,blue,opacity=0.5,mark size=1.5pt] table [x=tau, y=pos_zs, col sep=comma]{csv/vcas_pos_10.txt};

\addplot[mark=square*,orange,opacity=0.5,mark size=1.5pt] table [x=tau, y=pos_eq, col sep=comma]{csv/vcas_pos_-30.txt};
\addplot[mark=*,orange,opacity=0.5,mark size=1.5pt] table [x=tau, y=pos_zs, col sep=comma]{csv/vcas_pos_-30.txt};

\end{axis}
\end{tikzpicture}}}
\caption{Relative altitude $h$ of the two aircraft at time instants $k$ for equilibria  and zero-sum strategies for the VCAS[2] case study.}\label{fig:vcas_pos}
\vspace*{-0.2cm}
\end{figure}

\begin{figure}[!t]
\centering
\tikzstyle{state}=[rectangle,draw=blue,fill=blue!20,minimum size=6mm, inner sep = 0pt]
\tikzstyle{transition}=[circle, minimum size=0.05cm, fill=orange, inner sep = -1.5pt]
\begin{subfigure}{.49\textwidth}
\centering
\begin{tikzpicture}
        \node at (0,0) (sleg) [draw=blue,rectangle,fill=blue!20,label={above:{\small{state}}},label={right:{\small $s_k$}}] {\small \shortstack[c]{$(\mathit{tr}_{\mathit{own}},\mathit{ad}_{\mathit{own}})$ \\ $(\mathit{tr}_{\mathit{int}},\mathit{ad}_{\mathit{int}})$ \\ $(h,\dot{h}_{\mathit{own}},\dot{h}_{\mathit{int}},t)$}};
        
        \node at (0,-1.0) (tleg) [transition] {};

        \path[->]{
            (sleg) edge node[right,yshift=-1mm] {\small $(a_{\mathit{own}},a_{\mathit{int}})$} (tleg)
        };



        \node at (0,-2.0) (s0) [state,label={right:{\small $s_0$}}] {\small \shortstack[c]{$(4,1)$ \\ $(4,1)$ \\ $(50,-5,5,3)$}};
        \node at (0,-4.0) (s1) [state,label={right:{\small $s_1$}}] {\small \shortstack[c]{$(4,4)$ \\ $(4,1)$ \\ $(66,-14,8,2)$}};
        \node at (0,-6.0) (s2) [state,label={right:{\small $s_2$}}] {\small \shortstack[c]{$(4,1)$ \\ $(4,1)$ \\ $(91,-17,11,1)$}};
        \node at (0,-8.0) (s3) [state,label={right:{\small $s_3$}}] {\small \shortstack[c]{$(4,1)$ \\ $(4,1)$ \\ $(123,-20,14,0)$}}; 
        
        \node at (0,-3.0) (t0) [transition] {};
        \node at (0,-5.0) (t1) [transition] {};
        \node at (0,-7.0) (t2) [transition] {};
        
        
    
        
    \path[->]{
        (s0) edge node[right,yshift=-1mm] {\small $(-9.33,3)$} (t0)    
        (t0) edge node[left,yshift=0.5mm] {\small $1$} (s1)   
        (s1) edge node[right,yshift=-1mm] {\small $(-3,3)$} (t1)
        (t1) edge node[left,yshift=0.5mm] {\small $1$} (s2)
        (s2) edge node[right,yshift=-1mm] {\small $(-3,3)$} (t2)
        (t2) edge node[left,yshift=0.5mm] {\small $1$} (s3)
        };
\end{tikzpicture}
\end{subfigure}
\hfil
\begin{subfigure}{.49\textwidth}
\centering
\begin{tikzpicture}
        \node at (0,0) (s0) [state,label={right:{\small $s_0$}}] {\small \shortstack[c]{$(4,1)$ \\ $(4,1)$ \\ $(50,-5,5,4)$}} ;
        \node at (0,-2.0) (s1) [state,label={right:{\small $s_1$}}] {\small \shortstack[c]{$(4,4)$ \\ $(4,5)$ \\ $(68,-14,12,3)$}};
        \node at (0,-4.0) (s2) [state,label={right:{\small $s_2$}}] {\small \shortstack[c]{$(4,6)$ \\ $(4,7)$ \\ $(107,-26,24,2)$}};
        \node at (-1.25,-6.0) (s3) [state,label={left:{\small $s_3$}}] {\small \shortstack[c]{$(3,8)$ \\ $(4,1)$ \\ $(155,-26,21,1)$}}; 
        \node at (1.25,-6.0) (s4) [state,label={right:{\small $s_4$}}] {\small \shortstack[c]{$(4,8)$ \\ $(4,1)$ \\ $(155,-26,21,1)$}}; 
        \node at (-1.25,-8.0) (s5) [state,label={left:{\small $s_5$}}] {\small \shortstack[c]{$(3,1)$ \\ $(4,1)$ \\ $(199,-23,18,0)$}}; 
        \node at (1.25,-8.0) (s6)   [state,label={right:{\small $s_6$}}] {\small \shortstack[c]{$(4,1)$ \\ $(4,1)$ \\ $(199,-23,18,0)$}}; 
        
        \node at (0,-1.0) (t0) [transition] {};
        \node at (0,-3.0) (t1) [transition] {};
        \node at (0,-5.0) (t2) [transition] {};
        \node at (-1.25,-7.0) (t3) [transition] {};
        \node at (1.25,-7.0) (t4) [transition] {};
        
        
        
        
    \path[->]{
        (s0) edge node[right,yshift=-1mm] {\small $(-9.33,7.33)$} (t0)    
        (t0) edge node[left,yshift=0.5mm] {\small $1$} (s1)   
        (s1) edge node[right,yshift=-1mm] {\small $(-11.7,11.7)$} (t1)
        (t1) edge node[left,yshift=0.5mm] {\small $1$} (s2)
        (s2) edge node[right,yshift=-1mm] {\small $(0,-3)$} (t2)
        (t2) edge node[left,xshift=-1mm,pos=0.3] {\small $0.9$} (s3)
        (t2) edge node[right,xshift=1mm,,pos=0.3] {\small $0.1$}  (s4)
        (s3) edge node[right,yshift=-1mm] {\small $(3,-3)$} (t3)
        (s4) edge node[right,yshift=-1mm] {\small $(3,-3)$} (t4)
        (t3) edge node[left,yshift=0.5mm] {\small $0.1$} (s5)
        (t3) edge node[right,pos=0.5,xshift=3.5mm,yshift=0.5mm] {\small $0.9$} (s6.north west)
        (t4) edge node[right,yshift=0.5mm] {\small $1$} (s6)
        };
\end{tikzpicture}
\end{subfigure}
\caption{Optimal strategies for the VCAS[2] case study over two different time horizons,
using initial $t$ values of 3 (left) and 4 (right).}
\label{fig:vcas-strategy}
\vspace*{-0.2cm}
\end{figure}
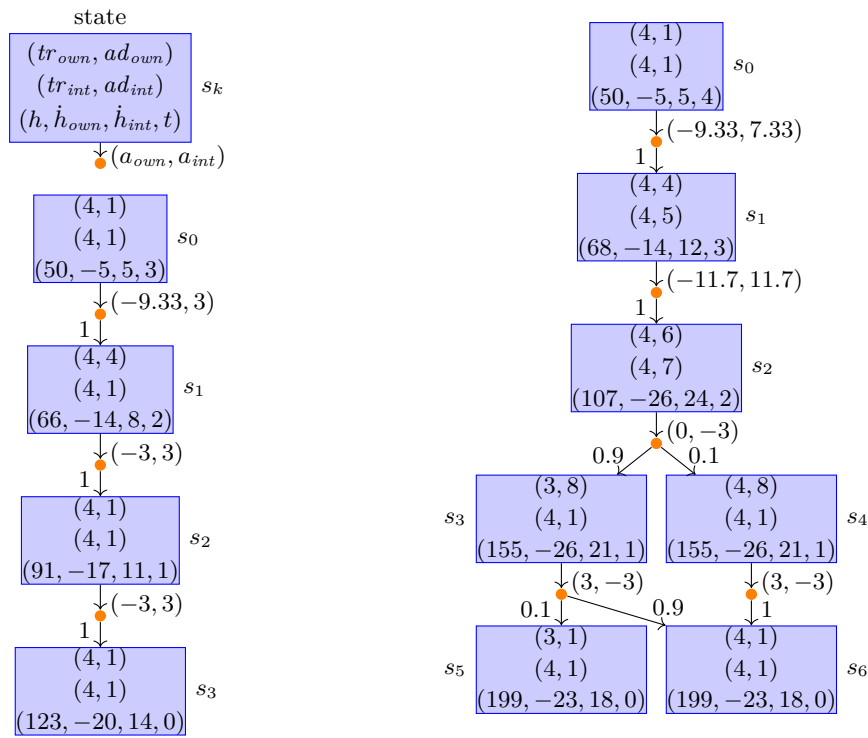

 
\begin{example}
The NS-CSG model of the VCAS[2] system described in \egref{vcas-example} is studied in \cite{YSD+22},
comparing equilibria strategies to the zero-sum strategies analysed in \cite{ABKL+20-2}.
\figref{fig:vcas_pos} plots the relative altitude $h$ of the two aircraft for equilibria  and zero-sum strategies when maximising this value for a given time instant $k$,
plotted for several different initial values of $h$.
It can be seen that, with respect to the safety criterion established in \cite{KDJ-MJK:19,ABKL+20-2},
i.e., avoiding an NMAC when two aircraft are separated by less than $100$ ft vertically (dotted line) and $500$ ft horizontally,
equilibria strategies allow the two aircraft to reach a safe configuration within a shorter horizon, which would be missed through a zero-sum analysis.

The analysis of \cite{YSD+22} also considers a reward structure that incorporates both the trust level and fuel consumption.
\figref{fig:vcas-strategy} shows the resulting equilibria strategies when both safety and trust are prioritised,
using two different time horizons. 
When $t=3$ initially (left), it is optimal to follow the advisories
and the trust levels $tr_{\mathit{own}}$ and $tr_{\mathit{int}}$ of the two aircraft
never decrease from their initial values of $4$.
However, when $t=3$ initially (right), the optimal strategy shows a deviation from the advisory,
denoted by action $a_{\mathit{own}}=0$, in state $s_2$, resulting in $tr_{\mathit{own}}$ dropping to $3$ in $s_3$ with probability $0.9$.
\end{example}

The experiments from \cite{YSD+22}, summarised above, and from \cite{nscsgs}
are developed using prototype tools that build upon parts of PRISM-games,
but full tool support for NS-CSGs is not yet provided.
Efforts in this direction are continuing.

\section{Conclusions and Future Challenges}

This paper has provided an overview of modelling, verification and strategy synthesis techniques that have been developed and implemented for concurrent stochastic games in the PRISM-games model checker. Through case studies, we have demonstrated the uses and advantages of zero-sum and equilibria-based reasoning in strategic decision making, highlighting Nash and correlated equlilibria in conjunction with two optimality criteria, social welfare and social fairness. We have also discussed recent trends in autonomous systems towards neural-symbolic architectures, and summarised the first steps towards developing a modelling framework to support the development of such AI-based systems. Despite some progress, many problems remain open in this area, in particular, the development of (efficient) approximate algorithms for (undiscounted infinite-horizon) temporal logic specifications where the underlying problem is undecidable even in the finite-state case~\cite{MHC03}.



\bibliography{bib}

\begin{thebibliography}{10}

\bibitem{ABKL+20}
M.~Akintunde, E.~Botoeva, P.~Kouvaros, and A.~Lomuscio.
\newblock Formal verification of neural agents in non-deterministic
  environments.
\newblock In {\em Proc.\ AAMAS'20}, pages 25--33. Springer, 2020.

\bibitem{ABKL+20-2}
M.~Akintunde, E.~Botoeva, P.~Kouvaros, and A.~Lomuscio.
\newblock {Verifying Strategic Abilities of Neural-symbolic Multi-agent
  Systems}.
\newblock In {\em Proc.\ KR'20}, pages 22--32. IJCAI Organization, 9 2020.

\bibitem{AH99}
R.~Alur and T.~Henzinger.
\newblock Reactive modules.
\newblock {\em Formal Methods in System Design}, 15(1):7--48, 1999.

\bibitem{alur2002alternating}
R.~Alur, T.~Henzinger, and O.~Kupferman.
\newblock Alternating-time temporal logic.
\newblock {\em Journal of the ACM}, 49(5):672--713, 2002.

\bibitem{AHM+98}
R.~Alur, T.~Henzinger, F.~Mang, S.~Qadeer, S.~Rajamani, and S.~Tasiran.
\newblock {MOCHA}: {M}odularity in model checking.
\newblock In {\em Proc. 10th Int. Conf. Computer Aided Verification (CAV'98)},
  volume 1427 of {\em LNCS}, pages 521--525, Vancouver, 1998. Springer.

\bibitem{Aum74}
R.~Aumann.
\newblock Subjectivity and correlation in randomized strategies.
\newblock {\em Journal of Mathematical Economics}, 1(1):67--96, 1974.

\bibitem{BMS14}
P.~Bouyer, N.~Markey, and D.~Stan.
\newblock Mixed {N}ash equilibria in concurrent games.
\newblock In {\em Proc.\ {FSTTCS}'14}, volume~29 of {\em {LIPICS}}, pages
  351--363, 2014.

\bibitem{BRE13}
R.~Brenguier.
\newblock {PRALINE}: A tool for computing {N}ash equilibria in concurrent
  games.
\newblock In {\em Proc.\ {CAV}'13}, volume 8044 of {\em LNCS}, pages 890--895.
  Springer, 2013.
\newblock \href{http://www.lsv.fr/Software/praline/}{lsv.fr/Software/praline/}.

\bibitem{BBG+19}
T.~Brihaye, V.~Bruy{\`e}re, A.~Goeminne, J.-F. Raskin, and M.~van~den Bogaard.
\newblock The complexity of subgame perfect equilibria in quantitative
  reachability games.
\newblock In {\em Proc.\ {CONCUR}'19}, volume 140 of {\em LIPIcs}, pages
  13:1--13:16. Leibniz-Zentrum f{\"u}r Informatik, 2019.

\bibitem{CLM+14}
P.~{\v{C}}erm{\'a}k, A.~Lomuscio, F.~Mogavero, and A.~Murano.
\newblock {MCMAS}-{SLK}: A model checker for the verification of strategy logic
  specifications.
\newblock In {\em Proc. {CAV}'14}, volume 8559 of {\em LNCS}, pages 525--532.
  Springer, 2014.

\bibitem{CAH13}
K.~Chatterjee, L.~de~Alfaro, and T.~Henzinger.
\newblock Strategy improvement for concurrent reachability and turn-based
  stochastic safety games.
\newblock {\em Journal of Computer and System Sciences}, 79(5):640--657, 2013.

\bibitem{CH08}
K.~Chatterjee and T.~Henzinger.
\newblock Value iteration.
\newblock In {\em 25 Years of Model Checking}, volume 5000 of {\em LNCS}, pages
  107--138. Springer, 2008.

\bibitem{CHJ+10}
K.~Chatterjee, T.~Henzinger, B.~Jobstmann, and A.~Radhakrishna.
\newblock {\sc Gist}: A solver for probabilistic games.
\newblock In {\em Proc.\ {CAV}'10}, volume 6174 of {\em LNCS}, pages 665--669.
  Springer, 2010.
\newblock \href{http://pub.ist.ac.at/gist/}{pub.ist.ac.at/gist/}.

\bibitem{CMJ04}
K.~Chatterjee, R.~Majumdar, and M.~Jurdzi\'nski.
\newblock On {N}ash equilibria in stochastic games.
\newblock In {\em Proc.\ {CSL}'04}, volume 3210 of {\em LNCS}, pages 26--40.
  Springer, 2004.

\bibitem{CFK+13b}
T.~Chen, V.~Forejt, M.~Kwiatkowska, D.~Parker, and A.~Simaitis.
\newblock Automatic verification of competitive stochastic systems.
\newblock {\em Formal Methods in System Design}, 43(1):61--92, 2013.

\bibitem{CKL+11}
C.~Cheng, A.~Knoll, M.~Luttenberger, and C.~Buckl.
\newblock {GAVS+}: An open platform for the research of algorithmic game
  solving.
\newblock In {\em Proc.\ {TACAS}'11}, volume 6605 of {\em LNCS}, pages
  258--261. Springer, 2011.
\newblock
  \href{https://sourceforge.net/projects/gavsplus/}{sourceforge.net/projects/gavsplus/}.

\bibitem{AHK07}
L.~de~Alfaro, T.~Henzinger, and O.~Kupferman.
\newblock Concurrent reachability games.
\newblock {\em Theoretical Computer Science}, 386(3):188--217, 2007.

\bibitem{AM04}
L.~de~Alfaro and R.~Majumdar.
\newblock Quantitative solution of omega-regular games.
\newblock {\em Journal of Computer and System Sciences}, 68(2):374--397, 2004.

\bibitem{Z3}
L.~De~Moura and N.~Bj{\o}rner.
\newblock Z3: An efficient {SMT} solver.
\newblock In {\em Proc.\ TACAS'08}, volume 4963 of {\em LNCS}, pages 337--340.
  Springer, 2008.
\newblock \href{https://github.com/Z3Prover/z3}{github.com/Z3Prover/z3}.

\bibitem{Dut14}
B.~Dutertre.
\newblock Yices 2.2.
\newblock In {\em Proc.\ {CAV}'14}, volume 8559 of {\em LNCS}, pages 737--744.
  Springer, 2014.
\newblock \href{http://yices.csl.sri.com}{yices.csl.sri.com}.

\bibitem{FKNP11}
V.~Forejt, M.~Kwiatkowska, G.~Norman, and D.~Parker.
\newblock Automated verification techniques for probabilistic systems.
\newblock In {\em SFM'11}, volume 6659 of {\em LNCS}, pages 53--113. Springer,
  2011.

\bibitem{gurobi}
{Gurobi Optimization, LLC}.
\newblock {Gurobi Optimizer Reference Manual}, 2021.
\newblock \href{https://www.gurobi.com}{gurobi.com}.

\bibitem{GNP+19}
J.~Gutierrez, M.~Najib, P.~Giuseppe, and M.~Wooldridge.
\newblock Equilibrium design for concurrent games.
\newblock In {\em Proc.\ {CONCUR}'19}, volume 140 of {\em LIPIcs}, pages
  22:1--22:16. Leibniz-Zentrum f{\"u}r Informatik, 2019.

\bibitem{GNP+18}
J.~Gutierrez, M.~Najib, G.~Perelli, and M.~Wooldridge.
\newblock {EVE}: A tool for temporal equilibrium analysis.
\newblock In {\em Proc. ATVA'18}, volume 11138 of {\em LNCS}, pages 551--557.
  Springer, 2018.
\newblock
  \href{https://github.com/eve-mas/eve-parity}{github.com/eve-mas/eve-parity}.

\bibitem{GLH+21}
Julian Gutierrez, Lewis Hammond, Anthony~W. Lin, Muhammad Najib, and Michael~J.
  Wooldridge.
\newblock Rational verification for probabilistic systems.
\newblock In {\em Proc. 18th International Conference on Principles of
  Knowledge Representation and Reasoning (KR'21)}, pages 312--322, 2021.

\bibitem{HJ94}
H.~Hansson and B.~Jonsson.
\newblock A logic for reasoning about time and reliability.
\newblock {\em FAC}, 6(5):512--535, 1994.

\bibitem{HR06}
L.~Hurwicz and S.~Reiter.
\newblock {\em Designing Economic Mechanisms}.
\newblock Cambridge University Press, 2006.

\bibitem{KDJ-MJK:19}
K.~Julian and M.~Kochenderfer.
\newblock A reachability method for verifying dynamical systems with deep
  neural network controllers.
\newblock {\em CoRR}, abs/1903.00520, 2019.

\bibitem{KDJ-SS-JBJ-MJK:19}
K.~Julian, S.~Sharma, J.{-}B. Jeannin, and M.~Kochenderfer.
\newblock Verifying aircraft collision avoidance neural networks through linear
  approximations of safe regions.
\newblock {\em CoRR}, abs/1903.00762, 2019.

\bibitem{Kar84}
N.~Karmarkar.
\newblock A new polynomial-time algorithm for linear programming.
\newblock {\em Combinatorica}, 4(4):373–395, 1984.

\bibitem{KSK76}
J.~Kemeny, J.~Snell, and A.~Knapp.
\newblock {\em Denumerable {M}arkov Chains}.
\newblock Springer, 1976.

\bibitem{KNPS20b}
M.~Kwiatkowska, G.~Norman, D.~Parker, and G.~Santos.
\newblock Multi-player equilibria verification for concurrent stochastic games.
\newblock In {\em Proc.\ QEST'20}, volume 12289 of {\em LNCS}, pages 74--95.
  Springer, 2020.

\bibitem{KNPS20}
M.~Kwiatkowska, G.~Norman, D.~Parker, and G.~Santos.
\newblock {PRISM}-games 3.0: Stochastic game verification with concurrency,
  equilibria and time.
\newblock In {\em Proc.\ CAV'20}, volume 12225 of {\em LNCS}, pages 475--487.
  Springer, 2020.

\bibitem{KNPS21}
M.~Kwiatkowska, G.~Norman, D.~Parker, and G.~Santos.
\newblock Automatic verification of concurrent stochastic systems.
\newblock {\em Formal Methods in System Design}, pages 1--63, 2021.

\bibitem{KNPS22}
M.~Kwiatkowska, G.~Norman, D.~Parker, and G.~Santos.
\newblock Correlated equilibria and fairness in concurrent stochastic games.
\newblock In {\em Proc.\ TACAS'22}, volume 13244 of {\em LNCS}, pages 60--78.
  Springer, 2022.

\bibitem{LaV00}
S.~LaValle.
\newblock Robot motion planning: A game-theoretic foundation.
\newblock {\em Algorithmica}, 26:430--465, 2000.

\bibitem{LH64}
C.~Lemke and Jr~J.~Howson.
\newblock Equilibrium points of bimatrix games.
\newblock {\em Journal of the Society for Industrial and Applied Mathematics},
  12(2):413--423, 1964.

\bibitem{LRTZ06}
M.~Littman, N.~Ravi, A.~Talwar, and M.~Zinkevich.
\newblock An efficient optimal-equilibrium algorithm for two-player game trees.
\newblock In {\em Proc.\ {UAI}'06}, pages 298--305. AUAI Press, 2006.

\bibitem{LQR09}
A.~Lomuscio, H.~Qu, and F.~Raimondi.
\newblock {MCMAS}: A model checker for the verification of multi-agent systems.
\newblock In {\em Proc. 21st Int. Conf. Computer Aided Verification (CAV'09)},
  volume 5643 of {\em LNCS}, pages 682--688. Springer, 2009.

\bibitem{LP17}
D.~Lozovanu and S.~Pickl.
\newblock Determining {N}ash equilibria for stochastic positional games with
  discounted payoffs.
\newblock In {\em Proc.\ {ADT}'17}, volume 10576 of {\em {LNAI}}, pages
  339--343. Springer, 2017.

\bibitem{LPS}
{LPSolve (version 5.5)}.
\newblock
  \href{http://lpsolve.sourceforge.net/5.5/}{lpsolve.sourceforge.net/5.5/}.

\bibitem{MHC03}
O.~Madani, S.~Hanks, and A.~Condon.
\newblock On the undecidability of probabilistic planning and related
  stochastic optimization problems.
\newblock {\em Artificial Intelligence}, 147(1--2):5--34, 2003.

\bibitem{MS18}
J.~Marden and J.~Shamma.
\newblock Game theory and control.
\newblock {\em Annual Review of Control, Robotics, and Autonomous Systems},
  1(1):105--134, 2018.

\bibitem{Mar98}
D.~Martin.
\newblock The determinacy of {B}lackwell games.
\newblock {\em J. Symbolic Logic}, 63(4):1565--1581, 1998.

\bibitem{GAMB}
R.~McKelvey, A.~McLennan, and T.~Turocy.
\newblock Gambit: Software tools for game theory.
\newblock \href{http://www.gambit-project.org}{gambit-project.org}.

\bibitem{NRTV07}
N.~Nisan, T.~Roughgarden, E.~Tardos, and V.~Vazirani.
\newblock {\em Algorithmic Game Theory}.
\newblock Cambridge University Press, 2007.

\bibitem{OR04}
M.~Osborne and A.~Rubinstein.
\newblock {\em An Introduction to Game Theory}.
\newblock Oxford University Press, 2004.

\bibitem{PNS04}
R.~Porter, E.~Nudelman, and Y.~Shoham.
\newblock Simple search methods for finding a {N}ash equilibrium.
\newblock In {\em Proc.\ AAAI'04}, pages 664--669. AAAI Press, 2004.

\bibitem{PPB15}
H.~Prasad, L.~Prashanth, and S.~Bhatnagar.
\newblock Two-timescale algorithms for learning {N}ash equilibria in
  general-sum stochastic games.
\newblock In {\em Proc.\ {AAMAS'15}}, pages 1371--1379. IFAAMAS, 2015.

\bibitem{EP14}
E.~Prisner.
\newblock {\em Game Theory Through Examples}.
\newblock Mathematical Association of America, 1 edition, 2014.

\bibitem{RF91}
T.~Raghavan and J.~Filar.
\newblock Algorithms for stochastic games --- a survey.
\newblock {\em Zeitschrift f{\"u}r Operations Research}, 35(6):437--472, Nov
  1991.

\bibitem{SGC05}
T.~Sandholm, A.~Gilpin, and V.~Conitzer.
\newblock Mixed-integer programming methods for finding {N}ash equilibria.
\newblock In {\em Proc.\ {AAAI}'05}, pages 495--501. AAAI Press, 2005.

\bibitem{SW+01}
U.~Schwalbe and P.~Walker.
\newblock Zermelo and the early history of game theory.
\newblock {\em Games and Economic Behavior}, 34(1):123--137, 2001.

\bibitem{Sha53}
L.~Shapley.
\newblock Stochastic games.
\newblock {\em PNAS}, 39:1095--1100, 1953.

\bibitem{TGW15}
A.~Toumi, J.~Gutierrez, and M.~Wooldridge.
\newblock A tool for the automated verification of {N}ash equilibria in
  concurrent games.
\newblock In {\em Proc.\ {ICTAC}'15}, volume 9399 of {\em LNCS}, pages
  583--594. Springer, 2015.

\bibitem{NEU28}
J.~von Neumann.
\newblock Zur theorie der gesellschaftsspiele.
\newblock {\em Mathematische Annalen}, 100:295--320, 1928.

\bibitem{NMK+44}
J.~von Neumann, O.~Morgenstern, H.~Kuhn, and A.~Rubinstein.
\newblock {\em Theory of Games and Economic Behavior}.
\newblock Princeton University Press, 1944.

\bibitem{Wac09}
A.~W{\"a}chter.
\newblock Short tutorial: Getting started with {\sc {i}popt} in 90 minutes.
\newblock In {\em Combinatorial Scientific Computing}, number 09061 in Dagstuhl
  Seminar Proceedings. Leibniz-Zentrum f{\"u}r Informatik, 2009.
\newblock \href{https://github.com/coin-or/Ipopt}{github.com/coin-or/Ipopt}.

\bibitem{YSD+22}
R.~Yan, G.~Santos, X.~Duan, D.~Parker, and M.~Kwiatkowska.
\newblock Finite-horizon equilibria for neuro-symbolic concurrent stochastic
  games.
\newblock In {\em Proc.\ UAI'22}, 2022.

\bibitem{nscsgs}
R.~Yan, G.~Santos, G.~Norman, D.~Parker, and M.~Kwiatkowska.
\newblock Strategy synthesis for zero-sum neuro-symbolic concurrent stochastic
  games.
\newblock \href{http://arxiv.org/abs/2202.06255 }{arXiv.2202.06255}, 2022.

\bibitem{pgwww}
{PRISM-games} web site.
\newblock
  \href{http://www.prismmodelchecker.org/games/}{prismmodelchecker.org/games/}.

\end{thebibliography}

\end{document}